\documentclass[preprint]{elsarticle} 

\usepackage[margin=2.7cm]{geometry} 

\usepackage{lineno,hyperref}
\modulolinenumbers[1]

\usepackage{subcaption}
\usepackage{amsmath}
\usepackage{amssymb}
\usepackage{color}
\usepackage[flushmargin,hang]{footmisc} 

\usepackage{comment}

\usepackage[euler]{textgreek}

\usepackage{enumitem}

\journal{Astroparticle Physics}

\begin{document}

\begin{frontmatter}

\title{Muon-induced neutrons in lead and copper at shallow depth}

\author[mymainaddress]{R. Knei\ss{}l\corref{mycorrespondingauthor}}\cortext[mycorrespondingauthor]{Corresponding author at: Max Planck Institute for Physics, 80805 Munich, Germany}\ead{raphael@mpp.mpg.de}

\author[mymainaddress]{A. Caldwell}

\author[mymainaddress,mysecondaryaddress]{Q. Du}

\author[mymainaddress,mythirdaddress]{A. Empl}

\author[mymainaddress]{C. Gooch}

\author[mymainaddress]{X. Liu}

\author[mymainaddress]{B. Majorovits}

\author[mymainaddress,myfourthaddress]{M. Palermo}

\author[mymainaddress]{O. Schulz}

\address[mymainaddress]{Max Planck Institute for Physics, 80805 Munich, Germany}

\address[mysecondaryaddress]{Present address: College of Physical Science and Technology, Sichuan University, 610064 Chengdu, China}

\address[mythirdaddress]{Department of Physics, University of Houston, 77204 Houston, USA}

\address[myfourthaddress]{Present address: Physics and Astronomy Department, University of Hawaii at Manoa, 96822 Honolulu, USA}

\clearpage

\begin{abstract}

Next generation low-background experiments require a detailed understanding of all possible radiation backgrounds. One important radiation source are muon-induced neutrons. Their production processes are up to now not fully understood. New measurements with MINIDEX (Muon-Induced Neutron Indirect Detection EXperiment) of the production of neutrons by cosmogenic muons in high-Z materials are reported. The setup is located at the T{\"u}bingen Shallow Underground Laboratory, which provides a vertical shielding depth of (13.2\,$\pm$\,0.8) meter water equivalent at the setup location. Muon-induced neutrons are identified by the detection of 2.2\,MeV gammas from their capture on hydrogen with high-purity germanium detectors.

The experimental results were compared to Geant4 Monte Carlo predictions. The measured rate of 2.2\,MeV neutron capture gammas for lead was found to be in good agreement with the Geant4 predicted rate. An external neutron yield of (7.2\,$^{\text{+\,0.7}}_{\text{$-$\,0.6}}$)\,$\cdot$\,10$^{-5}$\,g$^{-1}$\,cm$^{2}$ neutrons per tagged muon was determined for lead with the help of Geant4. For copper the measured rate was found to be a factor of 0.72\,$\pm$\,0.14 lower than the Geant4 predicted rate. Using this factor an external neutron yield of (2.1\,$\pm$\,0.4)\,$\cdot$\,10$^{-5}$\,g$^{-1}$\,cm$^{2}$ neutrons per tagged muon was obtained for copper.

An additional simulation was performed using the FLUKA Monte Carlo code. The FLUKA predicted rate of detected 2.2\,MeV neutron capture gammas for lead was also found to be in good agreement with the experimental value. A detailed comparison of muon interactions and neutron production in lead for Geant4 and FLUKA revealed large discrepancies in the description of photo-nuclear and muon-nuclear inelastic scattering reactions for muon energies at shallow underground sites. These results suggest that Geant4, when used with Geant4 recommended or standard physics lists, underpredicts the neutron production in photo-nuclear inelastic scattering reactions while at the same time it overpredicts the neutron production in muon-nuclear inelastic scattering reactions.

\end{abstract}


\end{frontmatter}

\section{Introduction and Motivation}

The next generation of low-background experiments need significantly reduced radiation backgrounds compared to current ones. One of the most critical background sources are muon-induced neutrons. These can either generate prompt signals in detectors or produce long lived radioactive isotopes by capture or inelastic scattering reactions. The subsequent decay of these isotopes can lead to significant backgrounds in the detectors if the correlation to the corresponding muon is lost. Although neutrons from radioactivity (energies up to $\approx$\,10\,MeV) outnumber muon-induced neutrons (energies up to several GeV) by typically 2 to 3 orders of magnitude they can be more efficiently shielded due to their lower energies. In order to reduce muon-induced backgrounds many experiments are built in deep underground laboratories. This leads to a reduction of the muon flux by many orders of magnitude~\cite{PhysRevD.86.010001}. Furthermore, the deployment of active shielding muon detection systems permits to identify and reject most prompt muon-induced events~\cite{gerda_active_shielding}. To shield detectors from the radioactivity of the installed experimental components as well as from the surroundings (e.g. cavern rock or concrete) often high-Z materials are selected. Two materials commonly used to shield ambient radioactivity are lead and copper~\cite{Heusser}. As these are typically located close to the detectors they can act themselves as a source of muon-induced neutrons.

In order to optimise future experiments with regards to muon-induced neutron backgrounds, reliable Monte Carlo (MC) simulations are crucial. In Geant4~\cite{Agostinelli2003250} each version and chosen physics list\footnote{A physics list is a set of physics models which are used in the simulation to describe the passage of particles through matter.} may yield different results. Only a small number of measurements of muon-induced neutrons produced in high-Z materials are available, which can be used to evaluate and tune simulations tools~\cite{Reichhart:2013xkd,Kluck:2013xga}. The goal of the MINIDEX~\cite{matteos_paper} (Muon-Induced Neutron Indirect Detection EXperiment) project is to provide reliable experimental data sets of muon-induced neutrons for different high-Z materials which are commonly used in low-background experiments. Dedicated Geant4-based MaGe~\cite{Boswell:2010mr} simulations are used for the evaluation of the propagation of muons and muon-induced particles through the setup. The Monte Carlo predictions can be compared to the experimental results. Unless otherwise specified, the results presented were obtained with MaGe (using Geant4 10.3). In order to get a better understanding of the experimental results and the predictive power of Geant4, additional studies with the simulation program FLUKA~\cite{Battistoni2015,ferrari2005fluka} were carried out. To investigate possible differences between FLUKA and Geant4, the interactions of muons and the subsequent production of neutrons in lead was studied in detail.

The experimental setup, the analysis strategy and the dedicated simulations were significantly improved since the commissioning run (Run~1) of MINIDEX~\cite{matteos_paper}. Two further runs were carried out since then. Muon-induced neutrons in lead only (Run~2) and muon-induced neutrons in lead and copper simultaneously (Run~3) were studied. The results, which are presented in the following, supersede the results of the commissioning run.

\section{Experimental Setup and Working Principle}

The design objective of MINIDEX was to create a simple, flexible and compact setup which can be used to measure muon-induced neutrons in different materials. The setup including all detectors, target materials, support structures and electronics is $\approx$\,1\,m wide, $\approx$\,1.5\,m long and $\approx$\,1.5\,m high. It can be easily moved to different locations and the target material can be exchanged within one day. The experiment is presently located in the T{\"u}bingen Shallow Underground Laboratory. The reason for going to a shallow underground site is to shield the cosmogenic atmospheric neutron flux while still preserving a high enough muon flux ($\approx$\,65\,muons\,s$^{-1}$\,m$^{-2}$). After the assembly and the start of each MINIDEX run, data taking is remotely controlled and monitored via an online interface.

\begin{figure}[htbp] 
\begin{subfigure}[b]{0.55\textwidth} 
\centering\includegraphics[scale=0.21]{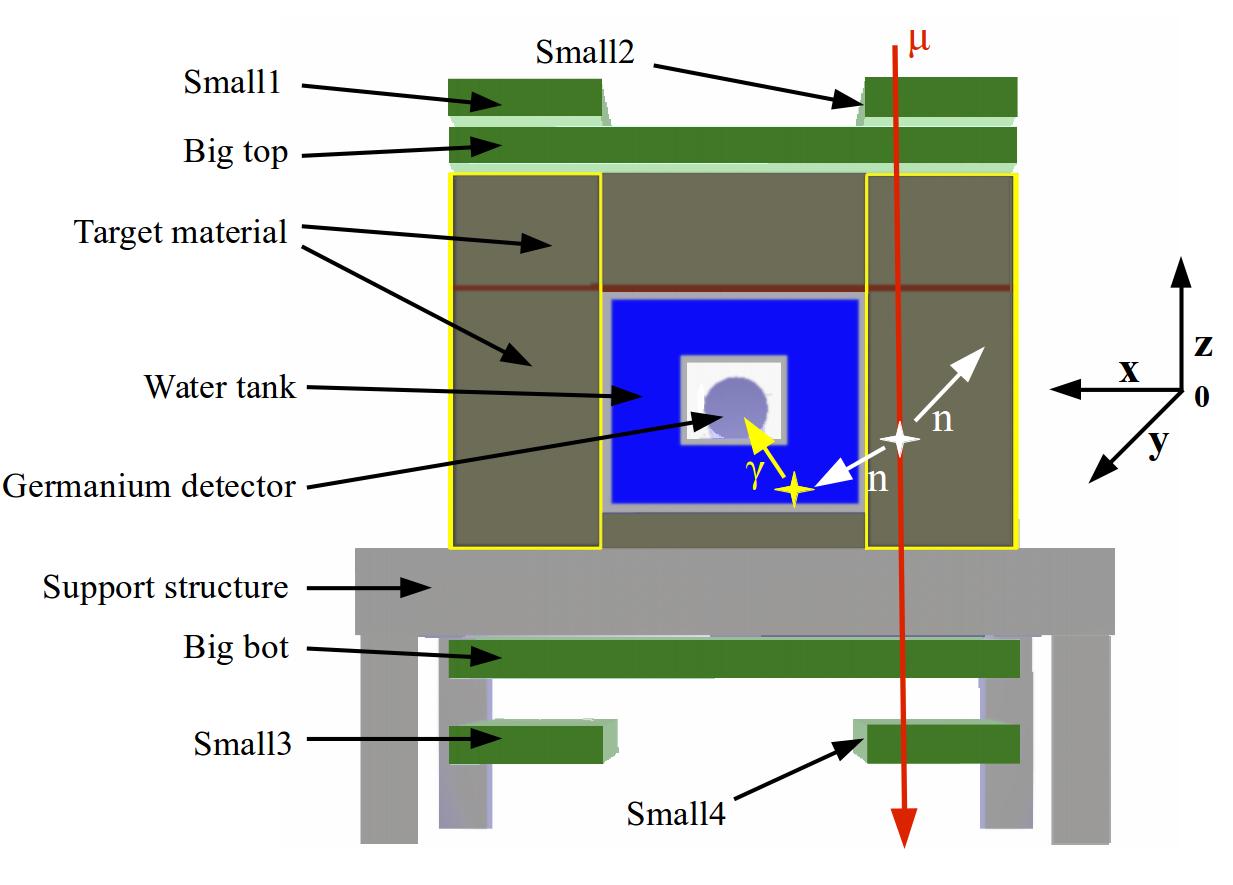}
\subcaption{}
\end{subfigure} 
\begin{subfigure}[b]{0.32\textwidth} 
\centering\includegraphics[scale=0.32]{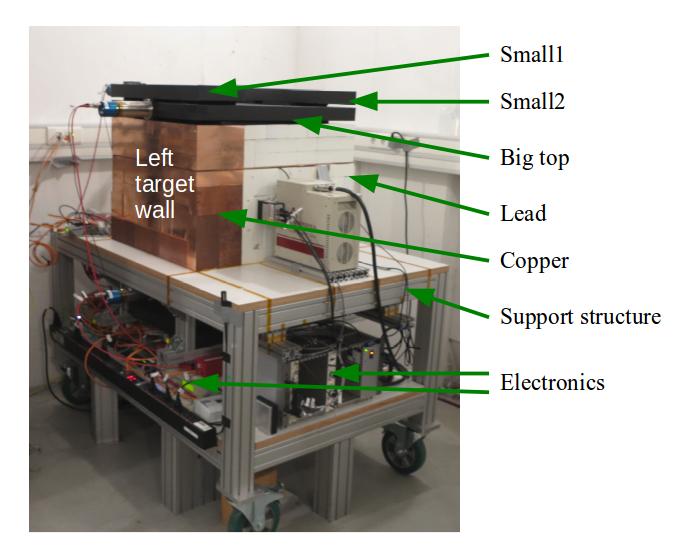}
\subcaption{}
\end{subfigure}
\caption{(a) Cross section of the MINIDEX Run~2 and Run~3 setup together with the working principle. The scintillation detectors (Big top, Big bot, Small1 - Small4), the target material, the water container and one of the germanium detectors are depicted. The target walls are indicated by yellow rectangles. Also a through-going muon, inducing neutrons and a neutron capture gamma, is indicated. In (b) a picture of the MINIDEX Run~3 setup is shown. Compared to Run~2 in which the whole target material consisted of lead, in Run~3 the left target wall was made of copper.} 
\label{fig:setup_working_principle_and_picture}
\end{figure}

A central cross section of the Run~2 and Run~3 MINIDEX setup with its support structure is shown in Fig.~\ref{fig:setup_working_principle_and_picture}(a).
MINIDEX consists of six scintillation detectors (Big top, Big bot, Small1 - Small4), the target material, a water filled plastic tank and two high-purity germanium detectors. The setup and the used electronic equipment are the same as presented in~\cite{matteos_paper} for the commissioning run of MINIDEX, except for the new muon tag system. In Fig.~\ref{fig:setup_working_principle_and_picture}(b) a picture of the Run~3 setup is shown. Compared to Run~2, in which the whole target material consisted of lead (density: 11.34\,g\,cm$^{-3}$), for Run~3 the left target wall was replaced with copper (density: 8.96\,g\,cm$^{-3}$).

\subsection{Working Principle and Signature of Muon-Induced Neutrons}

The working principle of measuring muon-induced neutrons with MINIDEX is depicted in Fig.~\ref{fig:setup_working_principle_and_picture}(a). A system of six scintillation detectors is used to identify muons traversing the setup. These muons, as well as muon-induced particles, can create neutrons~(n) in the target material of the setup. The neutrons propagate within the setup and some reach the water tank. In the water the neutrons thermalise and can get captured on hydrogen. Following this capture a single (2223.259\,$\pm$\,0.001)\,keV gamma~\cite{promt_gamma_database}, referred to as a 2.2\,MeV gamma, is emitted, which may be detected by one of the germanium detectors. 

Four small scintillation detectors permit to select muons that pass only through the target material, referred to as target walls. This is achieved by requesting a muon to pass through all four scintillation detectors on one side of the setup within a short time and deposit sufficient energy in each of them. Such events are denoted as muon tags. As an example, the depicted muon in Fig.~\ref{fig:setup_working_principle_and_picture}(a) represents a muon tag. After a muon tag, a signal from a delayed 2.2\,MeV capture gamma is searched for with the germanium detectors. Such an event, in the case of an identified 2.2\,MeV gamma within a predefined time window after a muon, is called a neutron signal. The details of this search have been determined by simulations and measurements.

\subsection{Muon Detection System}

The six plastic scintillation detectors of MINIDEX Run~2 and Run~3 are all 5\,cm thick. Four small scintillator panels of 20\,cm\,$\times$\,65\,cm and two big scintillator panels of 75\,cm\,$\times$\,65\,cm were used in the setup. A detailed description of the scintillation detectors and their positioning can be found in~\cite{Raphael_phd}. The scintillation detectors are made of BC-408 (Polyvinyltoluene: C$_{10}$H$_{11}$, density: 1.032\,g\,cm$^{-3}$)~\cite{Data_sheet_bc408} and are viewed by 2-inch diameter PMTs (ET Enterprises, type: 9266KFLB)~\cite{Data_sheet_PMT}. In comparison to the MINIDEX commissioning run where only two big scintillation detectors were used, Run~2 and Run~3 feature four additional smaller scintillation detectors. These are positioned above and below the target walls. The target walls are 20.0\,cm wide, 65.0\,cm long and 50.5\,cm high and are located on either side of the setup. In addition, the two big scintillation detectors, used in the commissioning run, were replaced because of their low muon detection efficiency ($\approx$\,87\,$\%$ and $\approx$\,93\,$\%$)~\cite{matteos_paper} with new ones. For all six newly installed scintillation detectors the efficiency to detect a through-going muon was determined to be $>$\,99\,$\%$~\cite{Raphael_phd}.

\subsection{Germanium Detectors}
\label{sec:germanium_detectors}

The two high-purity germanium detectors of MINIDEX are located in the centre of the setup, surrounded by the water tank. They are commercially available extended range coaxial high-purity germanium detectors from Mirion Technologies~\cite{Data_sheet_germanium}. Such detectors can be used for spectroscopy from 3\,keV up to energies above 10\,MeV~\cite{Data_sheet_extended_range_germanium_detector}. The two employed germanium crystals have a diameter of (69.0\,$\pm$\,0.5)\,mm/(70.5\,$\pm$\,0.5)\,mm, a length of (72.0\,$\pm$\,0.5)\,mm/(63.5\,$\pm$\,0.5)\,mm and a (45.0\,$\pm$\,0.5)\,mm/(37.5\,$\pm$\,0.5)\,mm deep borehole with a diameter of (9.5\,$\pm$\,0.5)\,mm in the centre. The crystals have a mass of (1.416\,$\pm$\,0.031)\,kg and (1.305\,$\pm$\,0.029)\,kg and a (1.0\,$\pm$\,0.5)\,mm thick dead layer on the n+ contact~\cite{Data_sheet_extended_range_germanium_detector}. The detectors are electrically cooled by dedicated devices which are located outside of the target material and can be seen in Fig.~\ref{fig:setup_working_principle_and_picture}(b).

\section{Data Taking and Data Selection}

\subsection{Energy Calibration and Detector Stability}

Both installed germanium detectors were calibrated individually every 10 hours. This was accomplished by exploiting two gamma lines from natural radioactivity (2.615\,MeV line from $^{208}$Tl and 1.461\,MeV line from $^{40}$K) originating mainly from materials in the vicinity of the detectors. At 1.461\,MeV and 2.615\,MeV the resolution of the germanium detectors (FWHM) was determined to be $\approx$\,2.1\,keV and $\approx$\,2.6\,keV, respectively. The resolution was monitored constantly over the whole data-taking period and no deterioration exceeding statistical uncertainty was observed.

The scintillation detectors were calibrated with the help of the muon spectra predicted by simulations of the Run~2 and Run~3 setups (see Section~\ref{sec:Simulation_Scintillator_Response_Muons}). The position of the most probable value of the Landau distribution for each scintillation detector was set to correspond to the value determined from simulation (all at $\approx$\,10\,MeV). The spectra of the scintillation detectors were monitored over the whole data-taking period. The observed shifts of the scintillation detector signal amplification were found to be below 0.2\,MeV.

\subsection{Triggering}

A flash ADC from Struck~\cite{Struck_ADC} was used to readout the scintillators and the high-purity germanium detectors. It operates with a 16 channel VME digitiser card and records traces with a 250\,MHz sampling rate and a 14\,bit resolution. A detailed description of the MINIDEX electronics can be found in~\cite{matteos_paper}. All scintillation and germanium detectors in the setup were independently triggered. For the germanium detectors energy depositions above a threshold of 20\,keV were recorded. A threshold of roughly 3\,MeV was selected for the scintillation detectors. For each energy deposition in the individual germanium detectors the amplitude of the charge signal was recorded together with its time stamp. Similarly, for each energy deposition in the individual scintillation detectors the amplitude of the current signal and the corresponding time stamp was recorded. All recorded energy depositions were stored and analysed offline.

\subsection{Muon Event Identification}
\label{sec:Muon_Tag_Determination}

The definition of a muon tag (also called tagged muon) requires coincident signals in four scintillation detectors on one side of the setup with an energy deposition $>$\,5\,MeV in each of them. The coincidence time window was set to $\pm$\,30\,ns with respect to the signal recorded by the Big top scintillation detector. These conditions eliminate accidental backgrounds, like energy depositions from radioactivity. From simulation it was determined that $>$\,95\,$\%$ of all muon tags result from events for which a muon passes through one target wall only. 

\begin{figure}[htbp]
\centering
\includegraphics[scale=0.45]{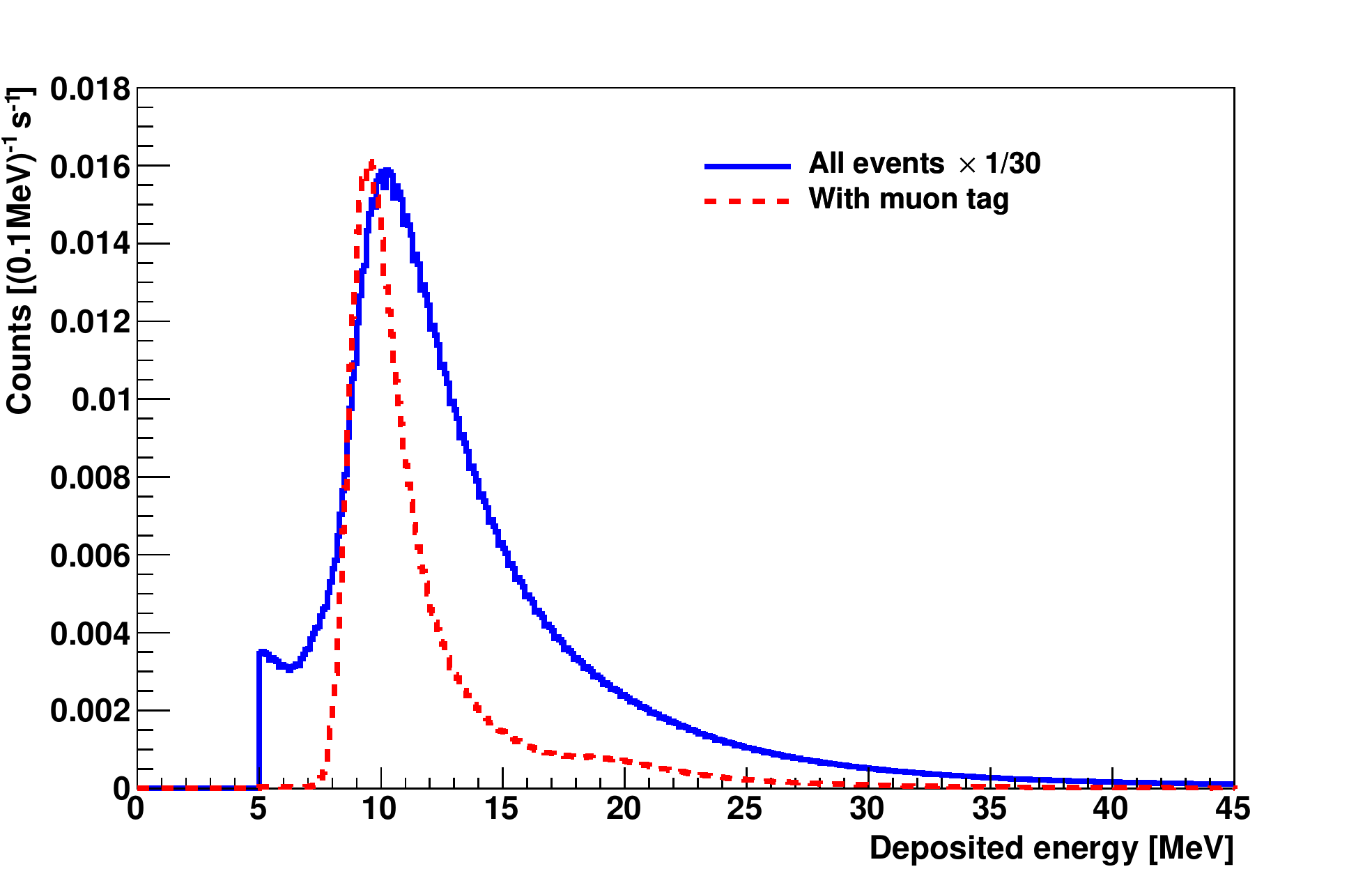} 
\caption{Measured energy spectrum of the Big top scintillation detector of MINIDEX Run~2 for two different sets of selected events. In solid blue all events with energies $>$\,5\,MeV (divided by a factor 30) are depicted. The spectrum in dashed red shows all events for the subset in which a muon tag on one side was found. Both spectra are normalised to the total lifetime of the run.}
\label{fig:big_top_all_events_muon_tag_events}
\end{figure}

In Fig.~\ref{fig:big_top_all_events_muon_tag_events} the measured energy spectrum of the Big top scintillation detector for Run~2 is depicted for energies $>$\,5\,MeV by the solid blue histogram (divided by a factor of 30).
Selecting only events with a muon tag results in the energy spectrum given by the dashed red histogram. In the latter, a sharpening of the Landau distribution as well as a shift to lower energies can be observed, resulting from the selection of muons by the geometrical acceptance of the tagging (maximum geometrically accepted angle with the z-axis: $\approx$\,55\,$^{\circ}$). Tag generating muons pass steeper through the scintillation detector than an average muon and, therefore, the mean deposited energy is smaller. At $\approx$\,20\,MeV a small bump can be observed in the spectrum. This bump results from the simultaneous passage of a muon and an additional coincident minimum ionising particle (mostly muon-induced electrons) through the scintillation detector.

\subsection{Experimental Data Sets}

MINIDEX Run~2 started in January 2016. Data were acquired until November 2016 which resulted in a data set with a lifetime of $\approx$\,260\,days. A total of 2.57\,$\cdot$\,10$^{7}$ muon tags have been identified for both muon tag sides, corresponding to a total muon tag rate of 1.14\,s$^{-1}$.  

After the exchange of the lead in the left target wall of the setup to copper, MINIDEX Run~3 data taking started. From November 2016 until May 2017 the Run~3 data set, corresponding to a lifetime of $\approx$\,166\,days, was acquired. For the lead side 8.23\,$\cdot$\,10$^{6}$ muon tags have been observed, leading to a muon tag rate of 0.57\,s$^{-1}$. For the copper side a number of 8.36\,$\cdot$\,10$^{6}$ muon tags and a corresponding muon tag rate of 0.58\,s$^{-1}$ was determined. The statistical uncertainty of all experimental muon tag rates is on the order of 10$^{-4}$\,s$^{-1}$.

\section{Predictions from Monte Carlo Simulations} 

To assess the signal detection efficiencies and the expected background of the measurement, the detector response to cosmogenic muons and muon-induced particles at the shallow underground site was simulated. The expected neutron signal rate in terms of detected 2.2\,MeV neutron capture gammas per muon tag was determined from the simulation. Further, the simulation was used to obtain an external neutron yield for lead and copper. The simulation was carried out in several steps. It starts with primary cosmic rays and a model of the earth's atmosphere and ends with the detector responses in MINIDEX. Each step utilises the output of the previous simulation step as input.

\subsection{Simulation of Cosmogenic Muons}
\label{sec:Simulation_Cosmogenic_Muons}

The FLUKA Monte Carlo code, version FLUKA 2011.2c, was used to provide the cosmogenic muon and muon-induced radiation field on a virtual sphere of 105\,cm radius, enclosing the MINIDEX experimental setup. The production of muons by primary cosmic rays at high altitude and their transport through the atmosphere to the surface of the earth was simulated with help of the FLUKA GCR tools, which are distributed with the standard FLUKA code~\cite{battistoni_2011}. The effect of the earth's magnetic field at the geographic location 48.5\,$^{\circ}$\,N, 9.1\,$^{\circ}$\,E as well as the elevation above sea level of 450\,m at the experimental site were taken into account.

\begin{figure}[htbp] 
\centering
\includegraphics[scale=0.48]{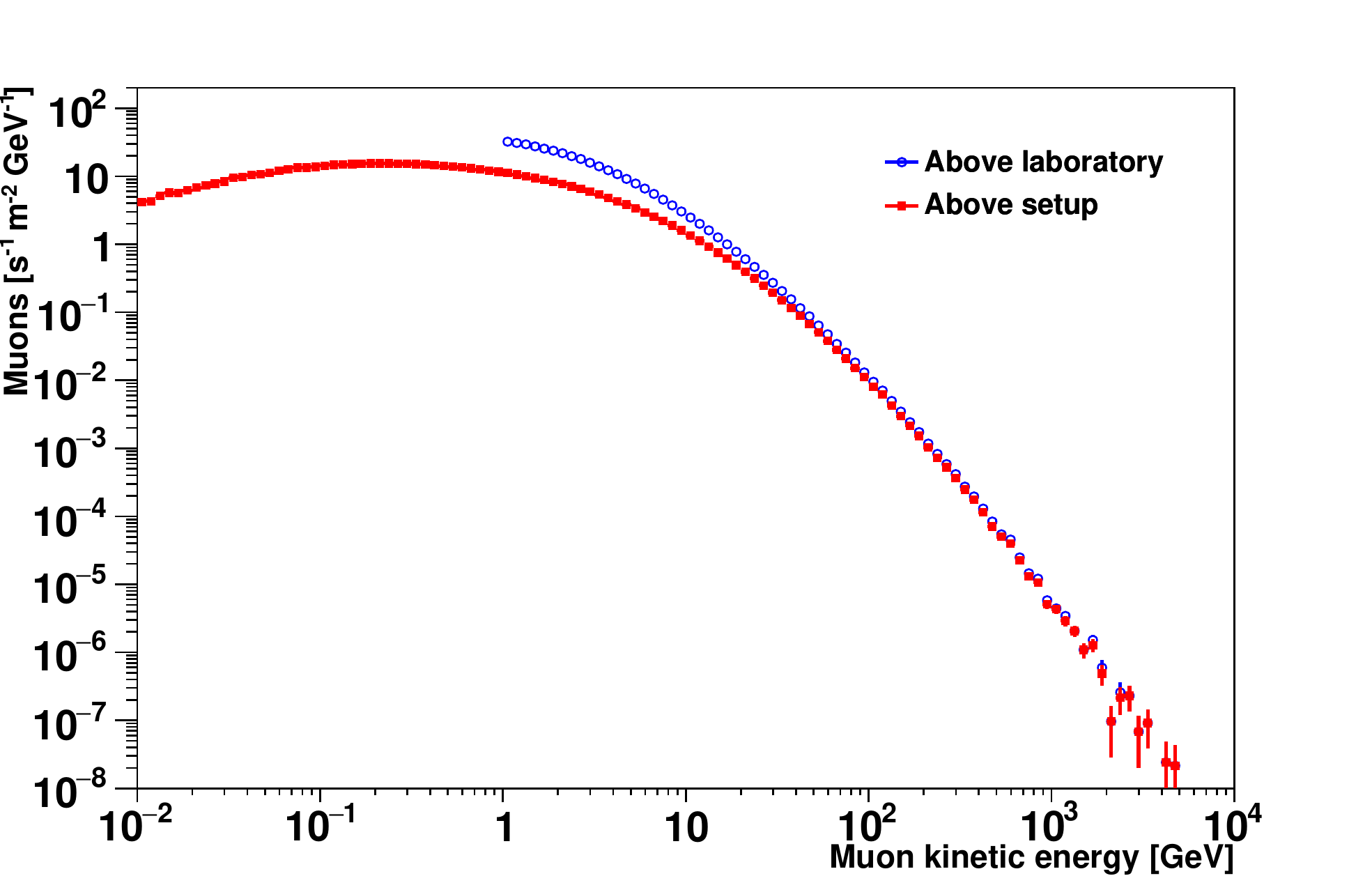}
\caption{Simulated energy spectrum of muons at the surface above the T\"ubingen Shallow Underground Laboratory (blue circles). In red squares the simulated energy spectrum of muons on a horizontal plane directly above the MINIDEX setup is given. These muons have a mean kinetic energy of (9.0\,$\pm$\,0.2)\,GeV. The spectra are normalised using an effective simulation lifetime as determined in Section~\ref{sec:Monte_Carlo_Data_Sets}. The displayed uncertainties are statistical only and are partially smaller than the marker size.}
\label{fig:Enegiers_Muons}
\end{figure}

\begin{figure}[htbp]
\centering
\includegraphics[scale=0.27]{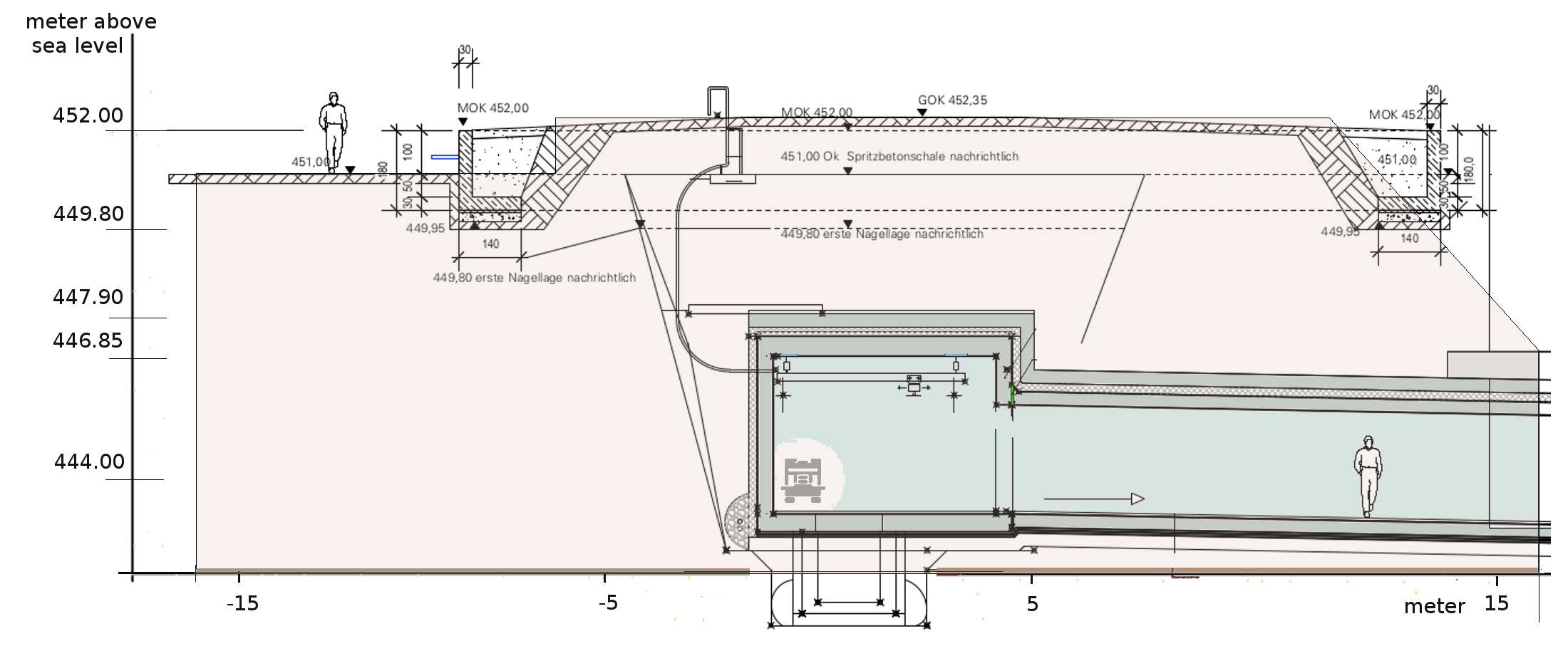}
\caption{Cross section of the underground laboratory and its surroundings at the University of T\"ubingen. The geometry implemented in FLUKA is shown by the coloured areas. The virtual sphere, on which all muons and muon-induced particles are recorded and subsequently removed from the simulation, is indicated. A detailed technical design drawing is superimposed for comparison~\cite{fritz}.}
\label{fig:LabFLUKA}
\end{figure}

The kinetic energy spectrum of muons reaching the surface of the earth is depicted by the blue histogram in Fig.~\ref{fig:Enegiers_Muons}.
The corresponding muon zenith angular distribution is well described by a cos$^2$ function. Subsequently, for all muons with kinetic energies above 1\,GeV, the propagation through the laboratory overburden was simulated\footnote{Only muons with kinetic energies considerably larger than 1\,GeV are able to reach the experimental setup.}.
For this purpose a model of the underground laboratory at the University of T\"ubingen and its surrounding was implemented in FLUKA. In Fig.~\ref{fig:LabFLUKA} a cross section of the implemented geometry is superimposed on a technical design drawing of the laboratory.

Information regarding the compressed soil composition above the laboratory indicates an average soil density of (2.2\,$\pm$\,0.2)\,g\,cm$^{-3}$ inside the overburden~\cite{goerz}. Together with additional structural material this corresponds to a vertical shielding depth of (13.2\,$\pm$\,0.8) meter water equivalent at the setup location. The simulated mean kinetic energy of muons reaching a horizontal plane directly above the experiment is (9.0\,$\pm$\,0.2)\,GeV (uncertainty resulting from density range). The kinetic energy distribution of these muons is displayed by the red histogram in Fig.~\ref{sec:Monte_Carlo_Data_Sets}. A report detailing the use of FLUKA in the context of MINIDEX and the construction of the muon and muon-induced radiation field at the underground laboratory is currently in preparation. 
 
On the order of 10$^{10}$ muons were started at the surface of the earth and propagated through the laboratory overburden. All muons and muon-induced particles entering the virtual sphere were recorded\footnote{The recorded information consists of the particle type, the location and momentum as well as the time.} and subsequently removed from the simulation. A data set consisting of $\approx$\,5\,$\cdot$\,10$^{9}$ events was prepared. These events, referred to as pre-recorded muon events, were available as input to the dedicated detector response simulations.

\subsection{Simulation of Scintillation Detector Response to Muons}
\label{sec:Simulation_Scintillator_Response_Muons}

The propagation of the particles from the pre-recorded muon events, located on the virtual sphere, through detailed geometrical models of the MINIDEX Run~2 and Run~3 setups was simulated with Geant4 (MaGe version: 17th February 2017 using Geant4 10.3 with Default MaGe physics list). Note that unless otherwise specified, the simulation results presented within this paper were obtained with Geant4. The pre-recorded muon events were simulated multiple times, using different random number generator starting seeds. On the order of 10$^{10}$ muon events were simulated for Run~2 and Run~3, respectively. For each run configuration a dedicated Monte Carlo data set was generated, recording the predicted energy depositions and the corresponding time stamps in the individual detectors.

Selecting events with energy depositions $>$\,5\,MeV in multiple scintillation detectors within a narrow $\pm$\,30\,ns time window defines different event classes depending on which scintillation detectors are considered. One example for a class of events is the muon tag (see Section~\ref{sec:Muon_Tag_Determination}) for which coincident signals of four scintillation detectors, either on the left or the right side of the setup, are required. The event rate and the shape of the obtained energy spectra depend on the chosen event class. The normalisation of the simulated rates is provided by the MC lifetime, which is obtained by relating the number of simulated coincidences between the Small2 and the Big top scintillation detector to the corresponding measured rate (see Section~\ref{sec:Monte_Carlo_Data_Sets}). A comparison of rates and spectra, determined for different event classes, permits to benchmark the predicted muon radiation field at the underground laboratory. The same energy cut and time window that was used for the experimental data was applied to the MC data. Agreement within a few percent between measured and simulated rates was found for all investigated event classes. Furthermore, agreement between the shape of the corresponding energy spectra was observed.

\begin{figure}[htbp] 
\begin{subfigure}[b]{0.475\textwidth}
\centering\includegraphics[scale=0.42]{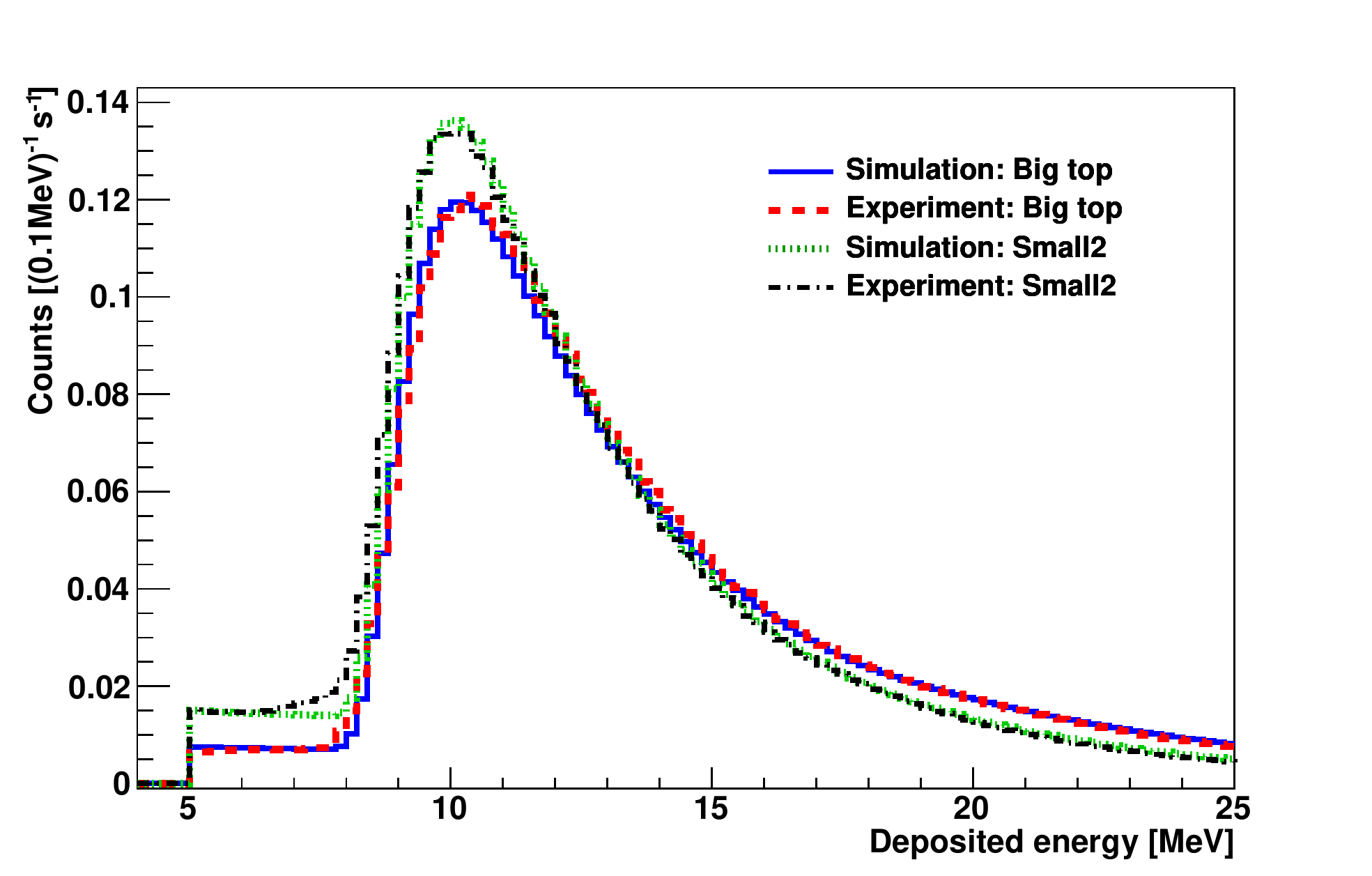}
\subcaption{}
\end{subfigure}
\begin{subfigure}[b]{0.56\textwidth} 
\centering\includegraphics[scale=0.42]{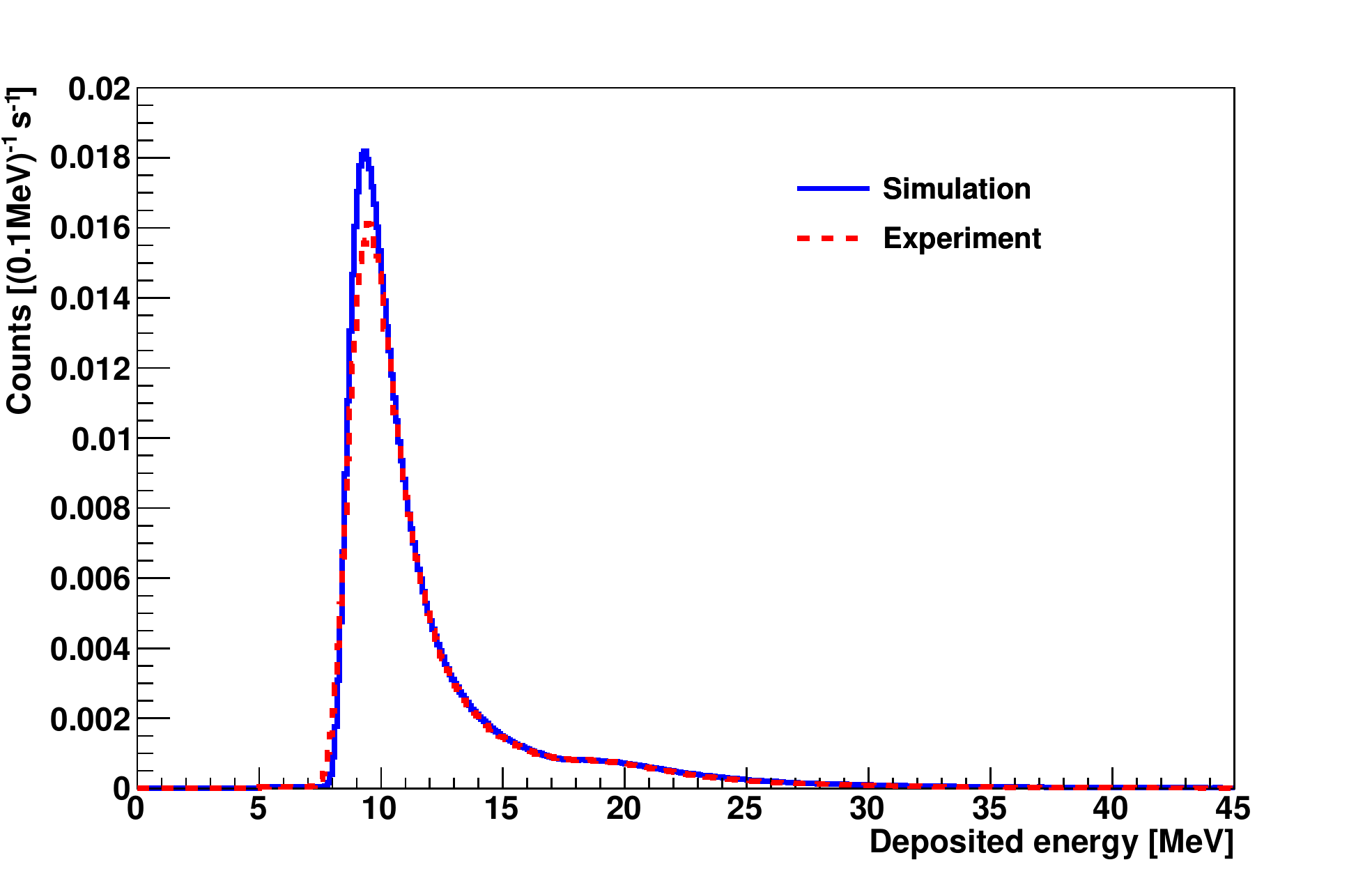}
\subcaption{}
\end{subfigure}
\caption{(a) Measured and simulated energy spectra of the Big top and the Small2 scintillation detectors in the case of coincident signals for MINIDEX Run~2 (for details see text). The measured spectra are normalised to the lifetime of the corresponding experimental data set. The spectra from the simulation are normalised with the help of (8.39\,$\pm$\,0.02)\,s$^{-1}$, the measured rate of these coincidences (see Section~\ref{sec:Monte_Carlo_Data_Sets}). In (b) the simulated and measured energy spectra of the Big top scintillation detector in the case of a found muon tag on one side are shown.}
\label{fig:Spec_MuonTag_NormCoinc}
\end{figure}

As an example of the agreement, the energy spectra for the class of events with a coincidence between the Big top and the Small2 scintillation detectors are shown in Fig.~\ref{fig:Spec_MuonTag_NormCoinc}(a). Here, the simulated rate of this event class was normalised to exactly match the measured rate. In Fig.~\ref{fig:Spec_MuonTag_NormCoinc}(b) the measured and the simulated energy spectra of the Big top scintillation detector, in the case of a muon tag on one side, are compared.
A difference of $\approx$\,6\,$\%$ was observed between predicted and measured rate. The mean kinetic energy of muons generating a muon tag, before entering the lead and copper target, was found to be (8.7\,$\pm$\,0.2)\,GeV and (8.5\,$\pm$\,0.2)\,GeV with a most probable energy of (1.05\,$\pm$\,0.05)\,GeV and (0.87\,$\pm$\,0.05)\,GeV (given uncertainties result from density range), respectively.

\subsection{Monte Carlo Data Sets} 
\label{sec:Monte_Carlo_Data_Sets}

The generated MC data set for MINIDEX Run~2 has a total of 2.12\,$\cdot$\,10$^{7}$ muon tags. For the MC data set corresponding to Run~3 a number of 1.69\,$\cdot$\,10$^{7}$/1.74\,$\cdot$\,10$^{7}$ muon tags were generated for the lead/copper side of the setup, respectively. In order to quantify the generated MC data sets corresponding to each run, an effective simulation lifetime was determined. This was carried out with the help of coincidences in the Small2 and the Big top scintillation detectors, shown in Fig.~\ref{fig:Spec_MuonTag_NormCoinc}(a) for MINIDEX Run~2. The rate of these coincidences in the experiment, R$_{\text{N}}^{\text{Exp}}$, is used to calculate the effective lifetimes of the generated MC data sets. A coincidence of two scintillation detectors was chosen, as the trigger rate of a single panel in the experiment contains lots of energy depositions from background sources. Furthermore, both chosen scintillation detectors are above the target material and are therefore (nearly) independent from interactions inside it. A value of (8.39\,$\pm$\,0.02)\,s$^{-1}$ was determined for R$_{\text{N}}^{\text{Exp}}$ for both runs. With this number an effective simulation lifetime of (202.9\,$\pm$\,0.5)\,days for the generated MC data set of Run~2 was obtained. For the MC data set corresponding to Run~3 an effective lifetime of (328.6\,$\pm$\,0.8)\,days was found. Note that the effective simulation lifetimes do not enter the following analysis.

\subsection{Neutron Signal Predictions}
\label{MC_Neutron_Signal_Predictions}

\begin{figure}[htbp] 
\centering      
\includegraphics[scale=0.48]{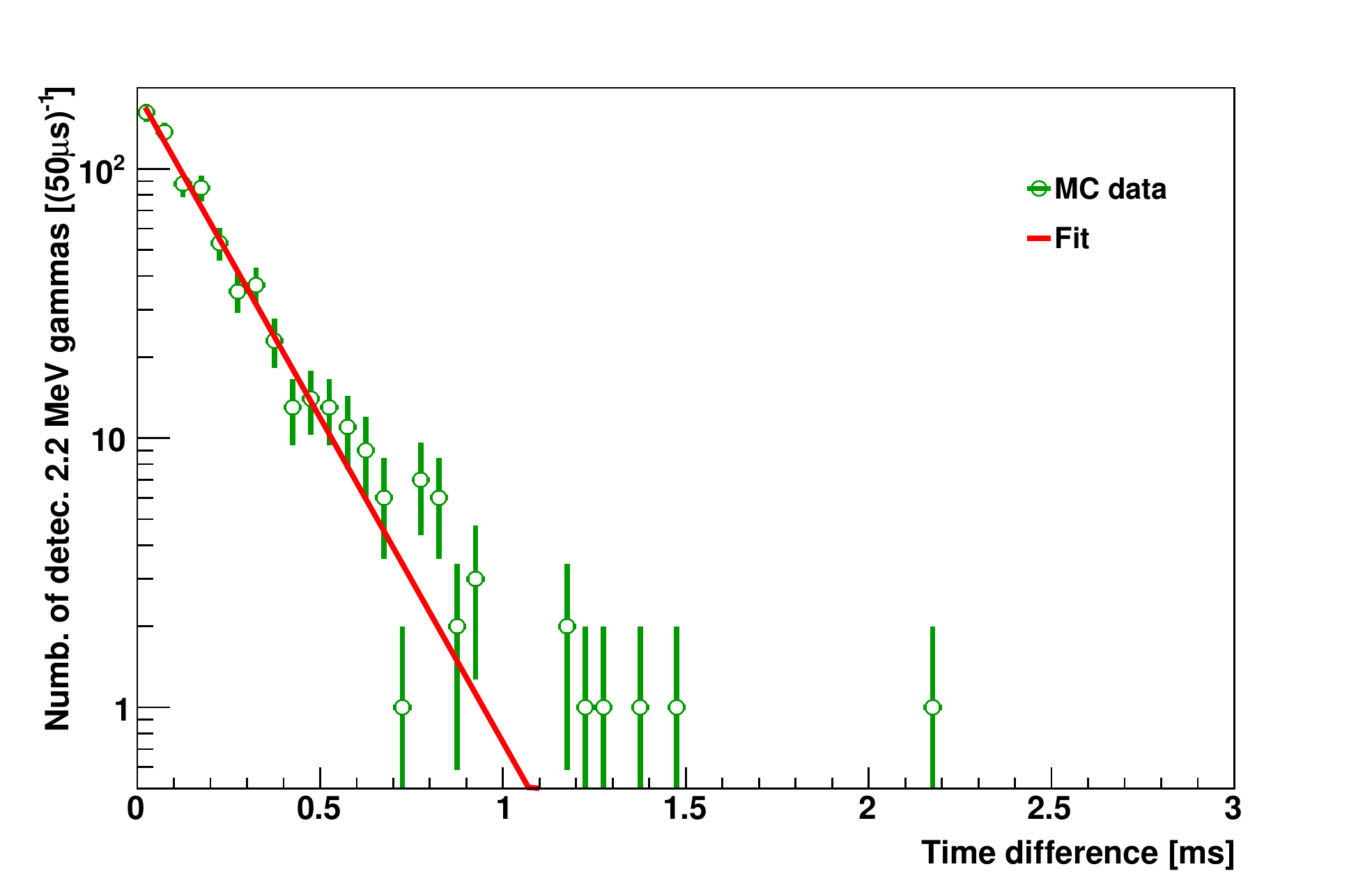}
\caption{Simulated time difference between an observed neutron signal in the germanium detectors and the correlated muon tag. The applied fit represents a falling exponential function. The mean delay obtained from the fit is (180\,$\pm$\,8)\,$\mu$s. The displayed uncertainties are statistical only.} 
\label{fig:MC_signal_fit}      
\end{figure}

The simulation was also used to determine the time distribution of the neutron signals after a muon tag. The obtained time distribution of the Run~2 MC data set is shown in Fig.~\ref{fig:MC_signal_fit}.
It can be seen that all neutron signals are expected within the first few ms after a muon tag. Overall (98.7\,$\pm$\,1.0)\,$\%$ of the neutron signals are predicted by the simulation to occur within 1\,ms after a muon tag. A fit, applied on the simulated time distribution of the neutrons signals, yields a mean delay of (180\,$\pm$\,8)\,$\mu$s. 

As no energy resolution is considered in the simulation, the number of neutron signals can be directly determined from the events found at 2224.37\,keV in the germanium detectors\footnote{The energy of gammas from neutron capture on hydrogen in the simulation is $\approx$\,1\,keV higher than the literature value of 2223.26\,keV~\cite{promt_gamma_database}. This bug is already known and was reported to the Geant4 collaboration~\cite{Abt2008}.}. However, when a realistic energy range corresponding to the $\pm$\,3\,$\sigma$ energy resolution of the germanium detectors at 2.2\,MeV is considered, a significant number of background events are obtained shortly after the muon tag. These originate from prompt charged particles which are induced by the muon during its passage through the setup. The simulation shows that by rejecting the first 7\,$\mu$s after a muon tag this background can be suppressed by a factor of $\approx$\,20, while (96.2\,$\pm$\,0.9)\,$\%$ of the neutron signal can be detected. In total, the simulation predicts that within a time window from 7\,$\mu$s to 1\,ms after a muon tag (94.9\,$\pm$\,1.3)\,$\%$ of all neutron signals can be detected. The 7\,$\mu$s and 1\,ms are chosen as time cuts in the experimental data analysis (see Section~\ref{sec:Extraction_Measured_Neutron_Signal}). The number of neutron signals per muon tag, determined for time differences $>$\,7\,$\mu$s in the simulation is denoted as R$_{\text{S}}^{\text{Sim}}$.

\subsection{External Neutron Yield Prediction}
\label{sec:Neutron_Yield_SIM}

The external neutron yield of an experiment, Y, quantifies the number of neutrons emerging from the surface of a target after the passage of a muon through it. As the number of emerging neutrons depends on the geometry of the target, Y is a setup specific quantity. In order to obtain a realistic neutron yield, discrepancies between the measured and the simulated observable need to be considered. Typically the neutron yield determined by simulation, Y$^{\text{Sim}}$, is scaled with F$_\text{S}$, the mismatch ratio between the measured and the simulated observable~\cite{Reichhart:2013xkd, Kluck:2013xga, lindote2009simulation}:
\begin{align}
\text{Y} =  \text{Y}^\text{Sim} \cdot \text{F}_\text{S} .
\label{eq:external_neutron_yield_1}
\end{align}
Y$^{\text{Sim}}$ is defined by
\begin{align}
\text{Y}^\text{Sim} = \frac{ \text{N}_\text{n}^\text{Sim} }{ \text{N}_\mu^\text{Sim} \cdot \overline{X}^\text{Sim}_\mu \cdot \rho_\text{target} } .
\label{eq:external_neutron_yield_2}
\end{align} 
N$_{\text{n}}^{\text{Sim}}$ represents the number of neutrons emerging in the simulation from the surface of the target walls towards the outside of the setup. N$_{\mu}^{\text{Sim}}$ stands for the number of muon tags, $\overline{X}^\text{Sim}_\mu$ represents the average track length of the simulated muons through the target walls of the setup and $\rho_{\text{target}}$ is the density of the target material. Only the corresponding surface, perpendicular to the x-axis (see Fig.~\ref{fig:setup_working_principle_and_picture}(a)), for the lead and copper side is considered in the following. Each neutron was recorded and subsequently removed from the simulation when leaving the setup through one of these two surfaces. With this approach the issue of correctly counting neutrons, passing through surfaces multiple times, is avoided. In the simulation, which makes use of the pre-recorded muon events (see Section~\ref{sec:Simulation_Cosmogenic_Muons}) and the implemented MINIDEX Run~3 setup, only events with a muon tag were considered. The used definition of the external neutron yield includes neutrons produced outside of the target walls. Furthermore, neutrons resulting from interactions of muon-induced particles within the target walls, that were generated outside the target walls, are included. The results for Y$^\text{Sim}$ together with the determined individual components of Equation~\ref{eq:external_neutron_yield_2} are shown in Table~\ref{tab:external_neutron_yield_SIM}.
\begin{table}[ht]
\centering
\caption{Results for Y$^\text{Sim}$ of lead and copper together with the individual quantities determined by a dedicated simulation of the MINIDEX Run~3 setup. Y$^\text{Sim}$ is calculated by using Equation~\ref{eq:external_neutron_yield_2}. The stated uncertainties are statistical only.}
\begin{tabular}{lll}
 \hline
  & \textbf{Lead} & \textbf{Copper}      \\
 \hline
  Y$^\text{Sim}$ [10$^{-5}$\,g$^{-1}$\,cm$^{2}$] neutrons per muon tag  & 6.96\,$\pm$\,0.02 & 2.93\,$\pm$\,0.02 \\
 \hline
 N$_{\mu}^{\text{Sim}}$  [10$^{6}$] & $\approx$\,1.97 & $\approx$\,2.05 \\
 N$_{\text{n}}^{\text{Sim}}$ & 82275 & 28135 \\
 $\overline{X}^\text{Sim}_\mu$ [cm] & 52.493\,$\pm$\,0.001 & 52.460\,$\pm$\,0.001 \\
 \hline
\end{tabular}
\label{tab:external_neutron_yield_SIM}
\end{table}
It was found that the total number of emitted neutrons per simulated muon tag from the corresponding surface is a factor of 2.37\,$\pm$\,0.02 higher for lead than for copper.

It is important to note that with the used procedure (described by Equation~\ref{eq:external_neutron_yield_1}) the number of all neutrons is scaled uniformly, independent of the neutron production processes. However, the quality of the description of the individual processes might be different for the simulation. Hints for this shortcoming are discussed in Section~\ref{sec:mage_fluka_comparison}. Combined with a significant energy dependent neutron detection efficiency~\cite{Raphael_phd}, this leads to the introduction of uncertainties which are very difficult to assess.

\section{Extraction of Measured Neutron Signal} 
\label{sec:Extraction_Measured_Neutron_Signal}

\begin{figure}[htbp]
\centering
\includegraphics[scale=0.4]{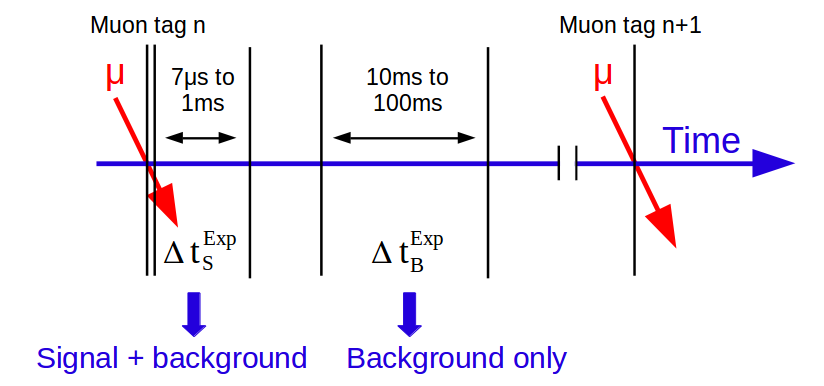}          
\caption{Schematic description of the analysis strategy applied in the experimental data analysis of MINIDEX Run~2 and Run~3.} 
\label{fig:time_windows}
\end{figure}

The analysis strategy to determine the number of neutron signals is illustrated in Fig.~\ref{fig:time_windows}.
After a muon tag a signal time window, $\Delta\text{t}_\text{S}^{\text{Exp}}$, from 7\,$\mu$s to 1\,ms is opened. In this time window neutron signals as well as background events are collected in the germanium detectors. From 10\,ms to 100\,ms a second time window, $\Delta\text{t}_\text{B}^{\text{Exp}}$, is opened in which only background events are collected. In the case of a muon tag and a subsequent second muon tag occurring within less than 100\,ms, the length of $\Delta\text{t}_\text{B}^{\text{Exp}}$ is shortened for the first muon tag accordingly. This adjustment leads to an average reduction of the length of $\Delta\text{t}_\text{B}^{\text{Exp}}$ by 9.4\,$\%$ in Run~2 and 3.2\,$\%$ in Run~3 (for both sides individually).

\begin{figure}[htbp]
\centering
\includegraphics[scale=0.48]{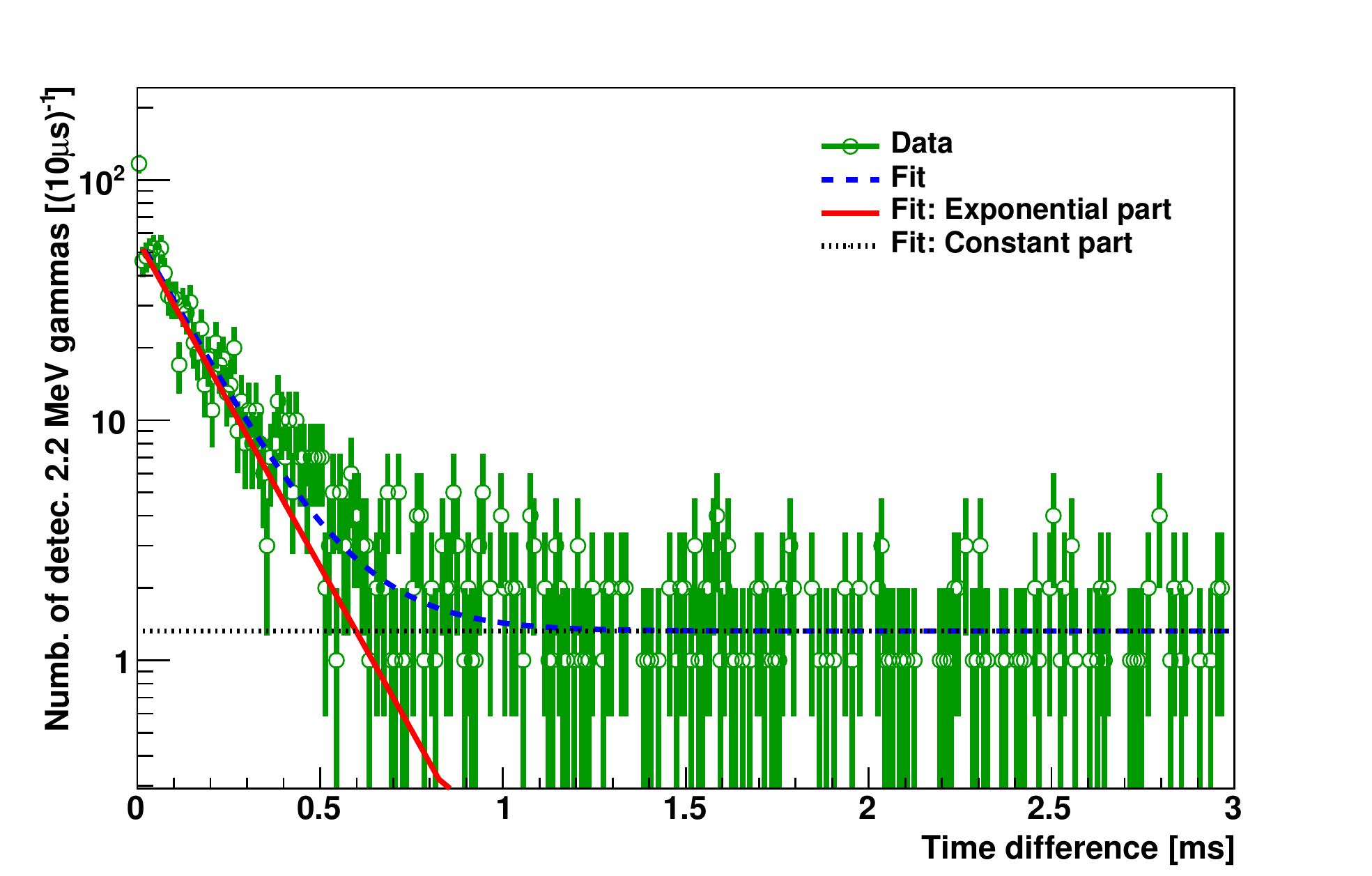}
\caption{Measured time difference between a muon tag and energy depositions in the germanium detectors within $\pm$\,3.7\,keV (corresponding to $\pm$\,3\,$\sigma$ energy resolution) around the energy of 2.2\,MeV capture gammas. The applied fit, depicted in dashed blue, consists of a falling exponential function (solid red) plus a constant (dotted black). The exponential function describes the neutron signals whereas the constant represents the background. The shown distribution was obtained for the MINIDEX Run~2 data set. The mean delay of the measured neutron signals, extracted from the exponential part of the fit, is (159\,$\pm$\,8)\,$\mu$s. The displayed uncertainties are statistical only.} 
\label{fig:mean_detection_time_fit}       
\end{figure}

Figure~\ref{fig:mean_detection_time_fit} shows the measured time difference between all events in the germanium detectors with an energy within (2223.3\,$\pm$\,3.7)\,keV ($\pm$\,3\,$\sigma$ of the energy resolution) and the previous muon tag, obtained for the Run~2 data set.
The number of detected events at 2.2\,MeV is clearly enhanced within $\approx$\,1\,ms after a muon tag, owing to muon-induced neutrons captured on hydrogen. This observation agrees well with the prediction from simulation (Fig.~\ref{fig:MC_signal_fit}). The measured time distribution was fit from 7\,$\mu$s to 3\,ms with the sum of a falling exponential function plus a constant. The falling exponential function describes the time evolution of the neutron signals whereas the constant function represents the muon tag independent background (constant over the measurement time). The fit yields a mean delay of (159\,$\pm$\,8)\,$\mu$s for the neutron signals, which is consistent within 2\,$\sigma$ uncertainties with the (180\,$\pm$\,8)\,$\mu$s predicted by simulation.

The lower time cut of 7\,$\mu$s in $\Delta\text{t}_\text{S}^{\text{Exp}}$ was chosen to reject the prompt background events as predicted by the simulation. These events can be seen in the first bin of the histogram in Fig.~\ref{fig:mean_detection_time_fit}. The spectrum is dominated by background events for times greater than 1\,ms after the muon tags. The time cuts of $\Delta\text{t}_\text{B}^{\text{Exp}}$ were chosen to be far away in time from the previous muon tag, and thus only non-muon tag related stochastic background events are collected (e.g. events from radioactivity or muons that did not generate a muon tag).

\begin{figure}[htbp] 
\begin{subfigure}[b]{0.475\textwidth}
\centering\includegraphics[scale=0.42]{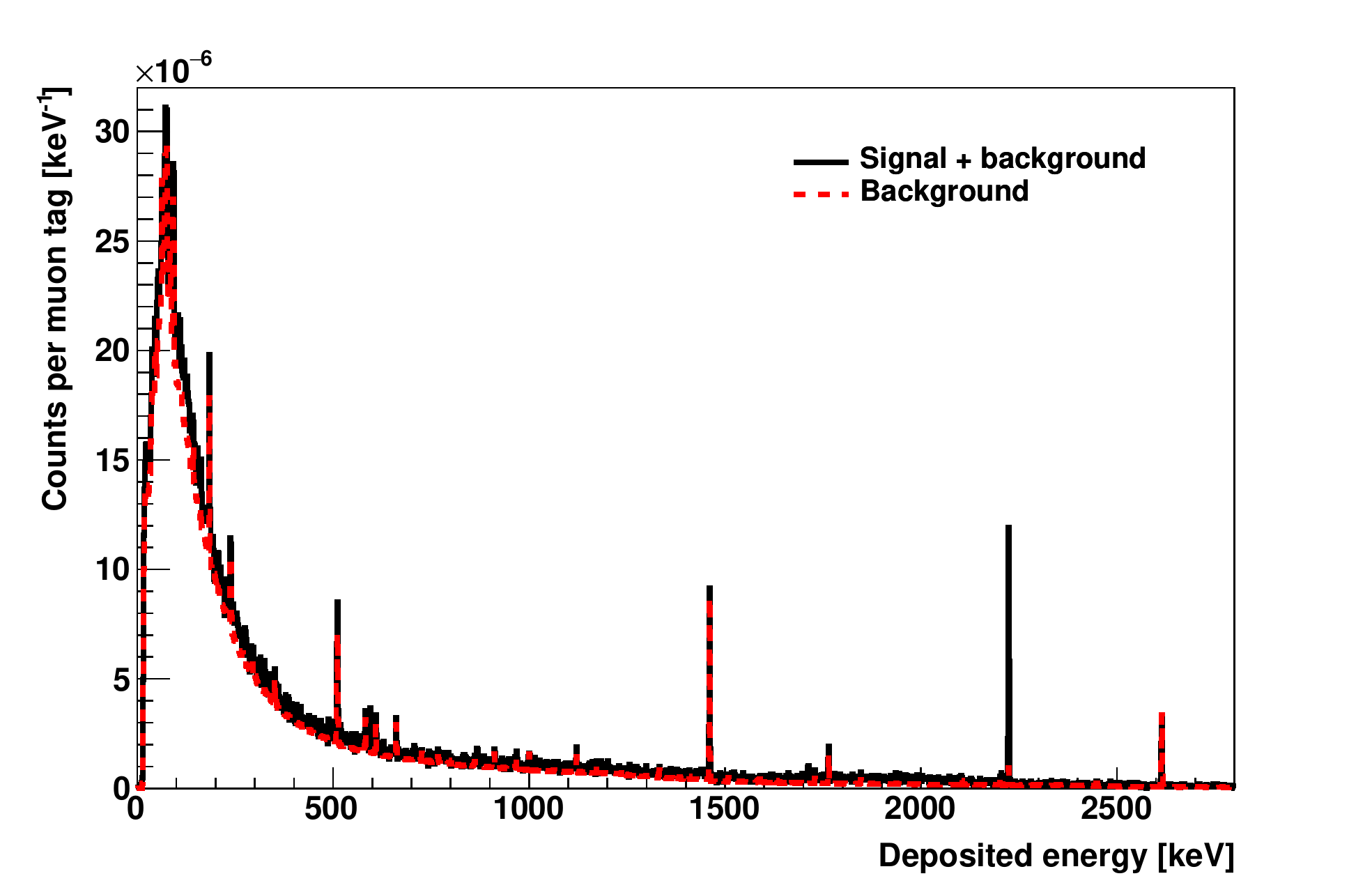}
\subcaption{}
\end{subfigure}
\begin{subfigure}[b]{0.56\textwidth} 
\centering\includegraphics[scale=0.42]{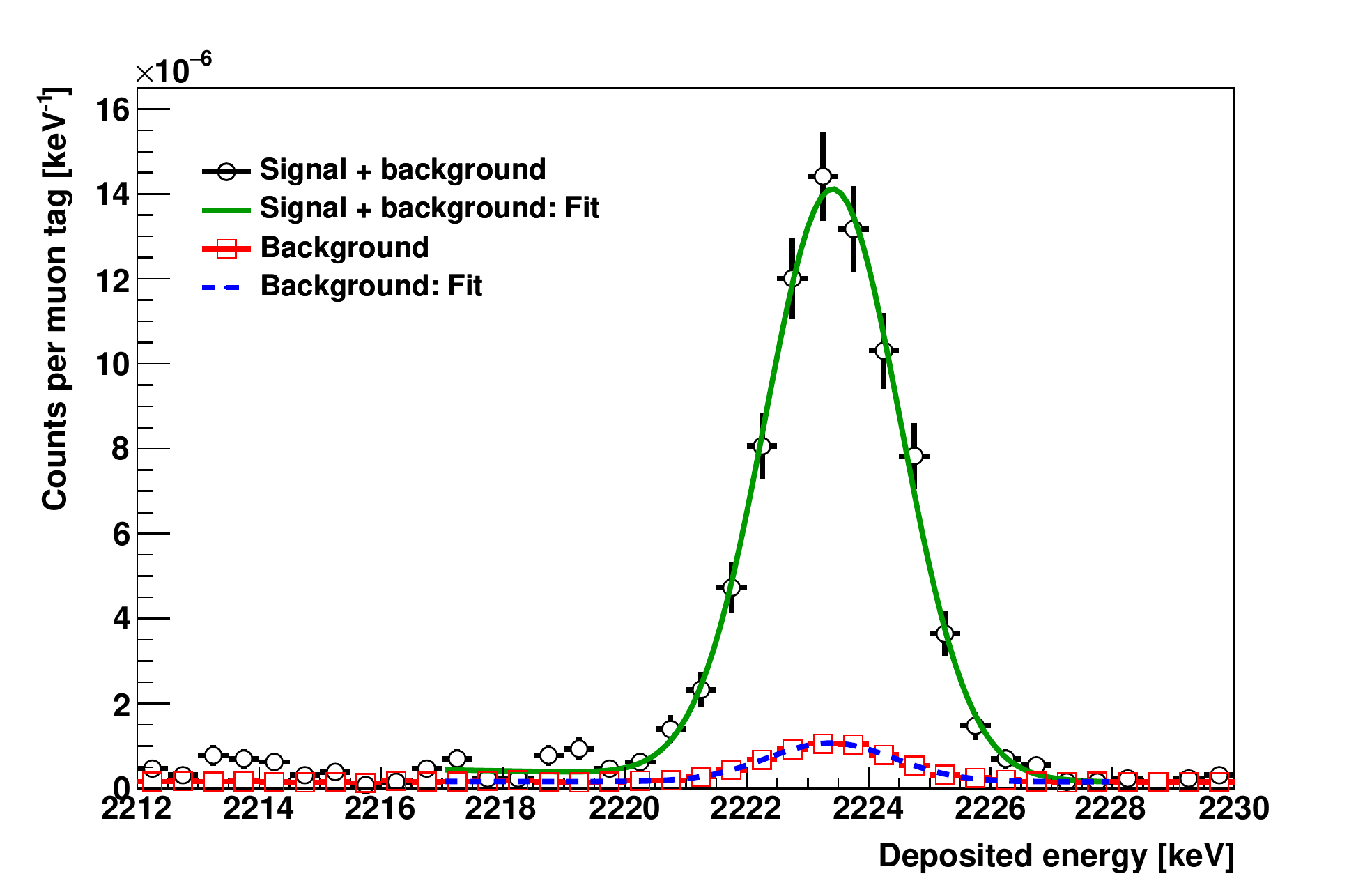}
\subcaption{}
\end{subfigure}
\caption{Spectra of the summed up time windows $\Delta\text{t}_{\text{S}}^{\text{Exp}}$ (signal plus background, solid black and black circles) and $\Delta\text{t}_{\text{B}}^{\text{Exp}}$ (background-only, dashed red and red squares) for the experimental data set of MINIDEX Run~2. In (a) all measured energies up to 2.8\,MeV and in (b) for a small energy region around the 2.2\,MeV neutron capture gamma peak (FWHM of the 2.2\,MeV gamma peak: $\approx$\,2.5\,keV) are shown. The signal plus background spectrum is normalised to the number of muons tags whereas the background-only spectrum is normalised to 993\,$\mu$s, the length of $\Delta\text{t}_{\text{S}}^{\text{Exp}}$. Gaussian plus first order polynomial function fits, applied to the 2.2\,MeV peaks, are depicted in (b) in solid green and dashed blue. These fits are used to determine R$_{\text{S}}^{\text{Exp}}$. The shown uncertainties in (b) are statistical only and are partially smaller than the marker size.}
\label{fig:Data_signal_fit}
\end{figure}

To extract R$_{\text{S}}^{\text{Exp}}$, the number of detected neutron signals per muon tag, all events measured in the germanium detectors after a muon tag within $\Delta\text{t}_{\text{S}}^{\text{Exp}}$ are summed up. The same is done for events within the time window $\Delta\text{t}_{\text{B}}^{\text{Exp}}$ to determine the number of stochastic 2.2\,MeV capture gamma background events per muon tag. The obtained spectra (signal plus background spectrum and background-only spectrum) for the data set of MINIDEX Run~2 are shown in Fig.~\ref{fig:Data_signal_fit}(a) for measured energies up to 2.8\,MeV and in Fig.~\ref{fig:Data_signal_fit}(b) for a small region around the energy of the 2.2\,MeV gammas.
The signal plus background spectrum is normalised to the number of found muon tags whereas the background-only spectrum is normalised to 993\,$\mu$s, the length of $\Delta\text{t}_{\text{S}}^{\text{Exp}}$. Both spectra are fit from 2.217\,MeV to 2.228\,MeV with a Gaussian plus a first order polynomial function (see Fig.~\ref{fig:Data_signal_fit}(b)). The areas of the Gaussian functions represent the number of detected 2.2\,MeV neutron capture gammas per muon tag. The underlying first order polynomial distributions describe stochastic background events in this energy range. R$_{\text{S}}^{\text{Exp}}$ is determined by subtracting the area of the Gaussian, determined for the normalised background-only spectrum, from the area of the Gaussian obtained for the normalised signal plus background spectrum.

\section{Results} 
\label{sec:Results}

In order to compare measured and simulated results of muon-induced neutrons, it is a real asset that MINIDEX measures the passage of muons through the setup itself. This provides the possibility to compare R$_{\text{S}}^{\text{Exp}}$ and R$_{\text{S}}^{\text{Sim}}$, independent of the absolute muon flux, as all values of R$_{\text{S}}$ are expressed in number of neutron signals per muon tag. It was found that the values of R$_{\text{S}}^{\text{Exp}}$ for lead, determined from the experimental Run~2 (both muon tag sides) and Run~3 (right muon tag side) data sets, are in agreement within statistical uncertainties. Further, it was observed that this also holds for R$_{\text{S}}^{\text{Sim}}$, obtained from the Run~2 and Run~3 MC data sets of lead. Hence, for lead the respective experimental and MC data sets of Run~2 and Run~3 were combined.

The analysis of the experimental lead data set yielded a total of 3.39\,$\cdot$\,10$^{7}$ muon tags and 1195\,$\pm$\,38 measured neutron signals (932\,$\pm$\,34 from Run~2 and 263\,$\pm$\,18 from Run~3). In the simulations a total
of 3.81\,$\cdot$\,10$^{7}$ muon tags and 
1285 neutron signals (712 from Run~2 and 573 from Run~3) were found. This results in R$_{\text{S}}^{\text{Exp}}$\,=\,(3.52\,$\pm$\,0.11)\,$\cdot$\,10$^{-5}$ and R$_{\text{S}}^{\text{Sim}}$\,=\,(3.37\,$\pm$\,0.09)\,$\cdot$\,10$^{-5}$ for lead.

For copper 8.36\,$\cdot$\,10$^{6}$ muon tags and 106\,$\pm$\,12 neutron signals were identified in the experimental data set. In contrast, the simulation yields 1.75\,$\cdot$\,10$^{7}$ muon tags and 309 neutron signals. This leads to R$_{\text{S}}^{\text{Exp}}$\,=\,(1.27\,$\pm$\,0.14)\,$\cdot$\,10$^{-5}$ and R$_{\text{S}}^{\text{Sim}}$\,=\,(1.77\,$\pm$\,0.10)\,$\cdot$\,10$^{-5}$ for copper.

\subsection{Investigation of Systematic Uncertainties}
\label{sec:syst_uncertain}

For the experimental and MC data analyses various sources of systematic uncertainties for R$_{\text{S}}$ were identified and investigated. In the following, all investigated sources are listed and the dominant ones are discussed in some detail. For each discussed contribution the maximal observed relative systematic uncertainty in the Run~2 or Run~3 data set is reported. A more detailed description of all investigated systematics can be found in~\cite{Raphael_phd} together with their individual contributions. \\

The five dominant sources of systematic uncertainties for R$_{\text{S}}^{\text{Exp}}$ of the experimental data analysis of the lead and copper data sets are:        \\
1. and 2. Dead layer thickness and size of the germanium crystals: The uncertainty on the dead layer thickness~\cite{Data_sheet_extended_range_germanium_detector} and the dimensions~\cite{Mail_mirion_technology_unceratainty_size_germanium} of the employed germanium crystals is specified by the producer to be 0.5\,mm. The dead layer thickness and crystal size is correlated to the active germanium detector mass and therefore to the 2.2\,MeV gamma detection probability. The determined relative systematic uncertainties for R$_{\text{S}}^{\text{Exp}}$ due to the uncertainties of the dead layer thickness and the size of the germanium crystals are $\pm$\,3\,$\%$ and $\pm$\,2\,$\%$, respectively.     \\
3. Energy reconstruction of measured pulses in the scintillation detectors, affecting the energy cuts: The accuracy of the reconstruction algorithm of energies deposited in the scintillation detectors depends on the rise time and the width of the recorded PMT pulses. For the reconstruction of the energy a trapezoidal filter with a gap time of 32\,ns and a peak time of 16\,ns was used. With the help of the simulation a maximal relative systematic uncertainty for R$_{\text{S}}^{\text{Exp}}$ of $\pm$\,2\,$\%$ was determined. \\
4. Scintillation detector signal gain stability and scintillation detector energy resolution: It was observed that slight signal gain changes in the scintillation detectors lead to shifts of the Landau distributions. These shifts were found to be always below $\pm$\,0.2\,MeV. Furthermore, upper limits on the energy resolution of the scintillation detectors were determined. A value of 1.2\,MeV/1.5\,MeV (sigma) was obtained for the small/big scintillation detectors, respectively~\cite{Raphael_phd}. The influence of both effects on R$_{\text{S}}^{\text{Exp}}$ was investigated by varying the energy cuts of the scintillation detectors applied in the muon tag determination procedure. The energy cuts of the small scintillation detectors were simultaneously increased/decreased by 1.4\,MeV whereas the ones of the big scintillation detectors by 1.7\,MeV and the analysis was repeated. A maximal relative systematic uncertainty for R$_{\text{S}}^{\text{Exp}}$ of $\pm$\,2\,$\%$ was obtained.     \\
5. Time stamps of energy depositions in germanium detectors: It was found that the time stamps of energy depositions in the germanium detectors are only precise within 1\,$\mu$s. The effect of this imprecision on the collection of 2.2\,MeV gammas in the time window $\Delta\text{t}_\text{S}^{\text{Exp}}$ was investigated. This was carried out by increasing and decreasing the lower time cut of 7\,$\mu$s, applied in the extraction of the measured neutron signals, by 1\,$\mu$s and repeating the data analysis. A maximal relative systematic uncertainty for R$_{\text{S}}^{\text{Exp}}$ of $\pm$\,2\,$\%$ was determined.

Further investigated systematic uncertainties for R$_{\text{S}}^{\text{Exp}}$ with contributions $\leq$\,1\,$\%$ are: accidental coincidences in the muon tag determination procedure, energy calibration of scintillation detectors, time stamps of energy depositions in scintillation detectors, function and energy range of signal and background determining fits, lower and higher time cut of $\Delta\text{t}_\text{S}^{\text{Exp}}$ and $\Delta\text{t}_{\text{B}}^{\text{Exp}}$, positioning of scintillation and germanium detectors, effect of changing lab equipment in the vicinity of the MINIDEX setup.    \\

Two sources of systematic uncertainties for R$_{\text{S}}^{\text{Sim}}$ of the MC data analysis for lead and copper have been investigated: \\
1. Physics list: Typically in Geant4, different physics lists can be used for the same purpose. To test the influence of the chosen physics list, the Run~2 setup was also simulated with the Shielding physics list~\cite{Shielding_Physics_List_webpage}. The Shielding physics list is recommended by the Geant4 collaboration for underground and low-background experiments as well as for neutron penetration studies. For the simulation the same pre-recorded muon events were used as an input (see Section~\ref{sec:Simulation_Cosmogenic_Muons}). A data set corresponding to an effective lifetime of $\approx$\,68\,days was generated and the same analysis procedure was applied as before. Within statistical uncertainties no change of R$_{\text{S}}^{\text{Sim}}$ with respect to the corresponding value obtained with the Default MaGe physics list was observed. \\
2. Soil density in the laboratory overburden:
The density of the soil in the overburden of the underground laboratory influences the energy spectrum, flux and composition of the particles reaching the setup. To study the influence on R$_{\text{S}}^{\text{Sim}}$, in addition to the pre-recorded muon events (see Section~\ref{sec:Simulation_Cosmogenic_Muons}) also muon events, generated for two further soil densities (density varied by $\pm$\,0.2\,g\,cm$^{-3}$), were simulated. MC data sets of the Run~2 setup with an effective lifetime of $\approx$\,80\,days were generated with Geant4 for both cases. The analysis of these data sets showed an increase/decrease of the mean muon energy (for tagged muons) by $\approx$\,0.2\,GeV for the higher/lower soil density, respectively. However, no systematic uncertainties for R$_{\text{S}}^{\text{Sim}}$ above statistical uncertainties were found.

No systematic uncertainties above statistical ones were found for either source. The systematic uncertainties for copper are also deduced from these investigations. Therefore, the statistical uncertainties of R$_{\text{S}}^{\text{Sim}}$ for lead and copper are adopted as a conservative estimate for the systematic uncertainties.

\subsection{Neutron Signal Rates}
\label{sec:Results_neutron_signals_rates}

The obtained neutron signal rates R$_{\text{S}}^{\text{Exp}}$ and R$_{\text{S}}^{\text{Sim}}$ of lead and copper with their systematic uncertainties are presented in Table~\ref{tab:results}.
\begin{table}[ht]
\centering
\caption{Measured and simulated neutron signal rates (R$_{\text{S}}^{\text{Exp}}$ and R$_{\text{S}}^{\text{Sim}}$) of lead and copper, determined from the corresponding experimental and MC data sets. The given neutron signal rates are expressed in number of neutron signals per muon tag. The simulated mean kinetic energy of tag generating muons in lead and copper, at the position before they enter the setup from the top, is (8.7\,$\pm$\,0.2)\,GeV and (8.5\,$\pm$\,0.2)\,GeV), respectively.}
\begin{tabular}{ll}
\hline
  & R$_{\text{S}}$ [10$^{-5}$]     \\
\hline
\textbf{Lead} &         \\
\hline
Experiment & 3.52\,$\pm$\,0.11\,(stat)\,$^{\text{+\,0.19}}_{\text{$-$\,0.14}}$\,(syst)  \\
Geant4  &  3.37\,$\pm$\,0.09\,(stat)\,$\pm$\,0.09\,(syst) \\
\hline
\textbf{Copper} &         \\
\hline
Experiment & 1.27\,$\pm$\,0.14\,(stat)\,$\pm$\,0.06\,(syst)  \\
Geant4     & 1.77\,$\pm$\,0.10\,(stat)\,$\pm$\,0.10\,(syst) \\
\hline
\end{tabular}
\label{tab:results}
\end{table}
The neutron signal rate predicted by the simulation for lead was found to be in good agreement with the corresponding experimental value. Using the obtained values of R$_{\text{S}}$, a factor of 1.04\,$^{\text{+\,0.10}}_{\text{$-$\,0.09}}$ (combined statistical and systematical uncertainty) was determined for the mismatch ratio F$_\text{S}^\text{Lead}$. For copper an overprediction of the neutron signal rate by the simulation was found, resulting in a value of 0.72\,$\pm$\,0.14 for F$_\text{S}^\text{Copp}$. It seems that Geant4 simulates lead better than copper. However, to cross-check the Geant4 predictions for lead, further investigations were carried out (see Section~\ref{sec:mage_fluka_comparison}).

\subsection{External Neutron Yield}
\label{sec:Neutron_Yield_Results}

\begin{figure}[htbp] 
\centering
\includegraphics[scale=0.48]{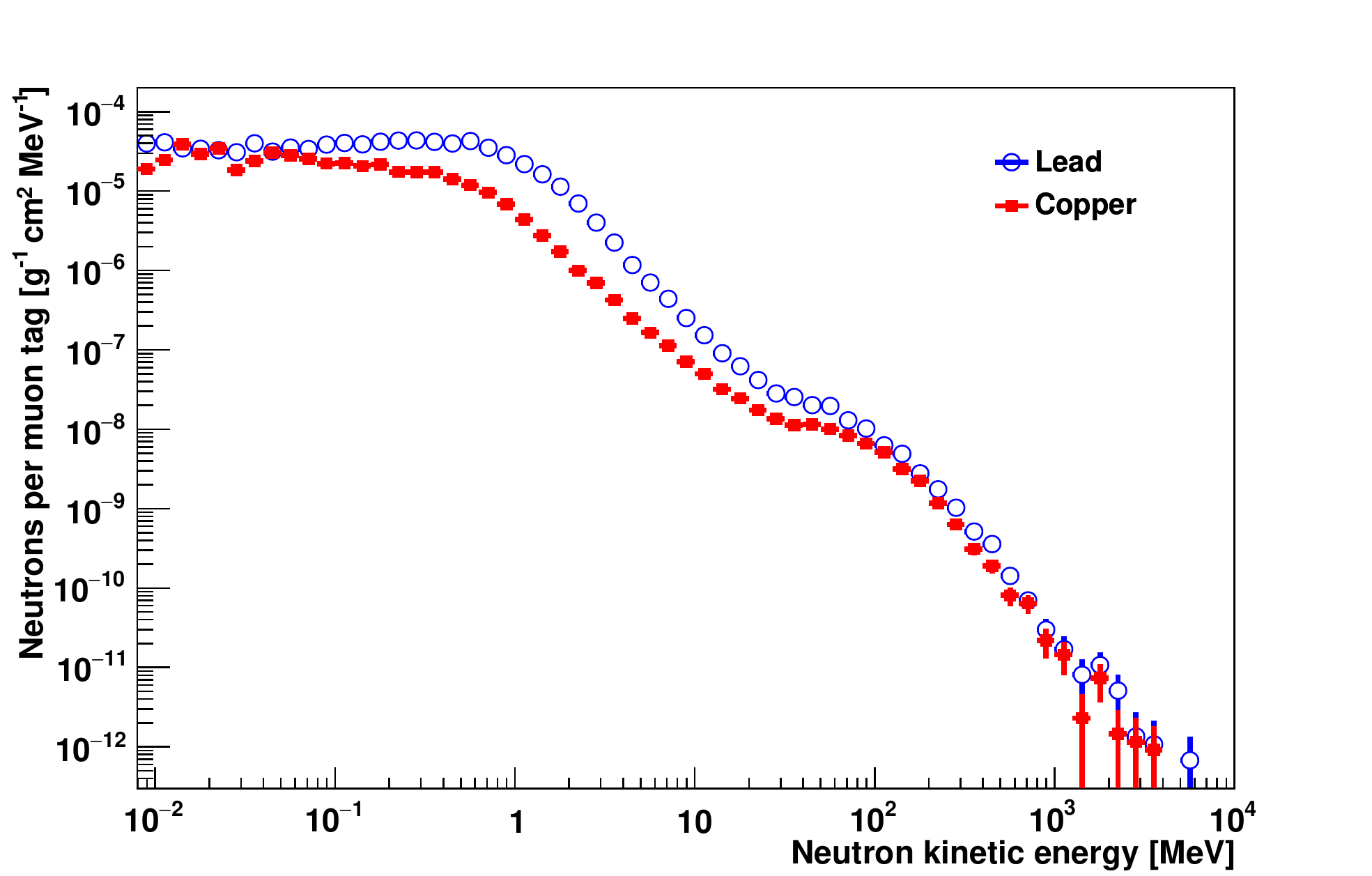}
\caption{Simulated energy spectra of muon-induced neutrons leaving the lead and copper surfaces of the MINIDEX Run~3 setup. Both spectra were scaled with the corresponding value of the mismatch ratio F$_\text{S}$ in order to represent the external neutron yield Y. The mean energy of the simulated muons for lead and copper, at the position before they enter the target from the top, is (8.7\,$\pm$\,0.2)\,GeV and (8.5\,$\pm$\,0.2)\,GeV, respectively. For the emitted neutrons a mean energy of (4.9\,$\pm$\,0.2)\,MeV and (8.9\,$\pm$\,0.3)\,MeV for lead and copper was found, respectively. The displayed  uncertainties are statistical only and are partially smaller than the marker size.} 
\label{fig:neutron_spectra_external_neutron_yield}
\end{figure}

With the determined values of Y$^\text{Sim}$ (from Section~\ref{sec:Neutron_Yield_SIM}) and F$_\text{S}$ (from Section~\ref{sec:Results_neutron_signals_rates}) Y was calculated for lead and copper using Equation~\ref{eq:external_neutron_yield_1}. The resulting values are given in Table~\ref{tab:external_neutron_yield_Results}.
\begin{table}[ht]
\centering
\caption{Determined values of the external neutron yield Y for lead and copper. Y is a factor of 3.4\,$^{\text{+\,0.8}}_{\text{$-$\,0.7}}$ higher for lead than for copper. The total uncertainties are reported.} 
\begin{tabular}{lll}
 \hline
         & \textbf{Lead} & \textbf{Copper}      \\
 \hline
  Y [10$^{-5}$\,g$^{-1}$\,cm$^{2}$] neutrons per muon tag  & 7.2\,$^{\text{+\,0.7}}_{\text{$-$\,0.6}}$ & 2.1\,$\pm$\,0.4    \\
 \hline
\end{tabular}
\label{tab:external_neutron_yield_Results}
\end{table}
For lead the total number of emitted neutrons per simulated muon tag is a factor of 3.4\,$^{\text{+\,0.8}}_{\text{$-$\,0.7}}$ larger than for copper. In Fig.~\ref{fig:neutron_spectra_external_neutron_yield} the simulated energy spectra of the muon-induced neutrons leaving the lead and copper surfaces are shown. The simulated spectra were scaled with the corresponding mismatch ratio F$_\text{S}$.
It can be seen that for nearly all neutron energies (up to a few hundred MeV) the number of emitted neutrons from the lead surface is higher than from the copper surface. The simulated mean energy of neutrons emitted from the lead surfaces is (4.9\,$\pm$\,0.2)\,MeV whereas for the copper surface it is (8.9\,$\pm$\,0.3)\,MeV.

\section{Geant4 and FLUKA Predictions for Muon-Induced Neutron Production in Lead} 
\label{sec:mage_fluka_comparison}

A simulation of the Run~2 setup, making use of the same pre-recorded muon events (see Section~\ref{sec:Simulation_Cosmogenic_Muons}), was carried out with the FLUKA Monte Carlo program.  The FLUKA predicted neutron signal rate R$_{\text{S}}^{\text{Sim}}$ for lead is (3.60\,$\pm$\,0.13\,(stat))\,$\cdot$\,10$^{-5}$ and agrees well, both with the experimental neutron signal rate R$_{\text{S}}^{\text{Exp}}$ and the value for R$_{\text{S}}^{\text{Sim}}$ obtained with Geant4.

Despite the agreement, a more detailed comparison of simulation predictions for the muon-induced neutron production in lead by the two programs was carried out. This comparison is motivated by recent publications~\cite{Quantification_validity_phonton_nuclear,Luo2016}, pointing out discrepancies between Geant4 predictions and experimental findings for the neutron production by photo-nuclear inelastic scattering reactions in high-Z materials (tungsten, gold and lead). At the same time, FLUKA predictions were reported to describe experimental observations reasonably well.

Only muon interactions inside the lead target walls for events with a muon tag were selected for this study. The predictions for the muon interactions and the resulting secondary particles are compared in Section~\ref{sec:Geant4_Fluka_Muon_Interaction}. All neutrons generated as a consequence of the initial muon interactions in the lead target walls were recorded at the point of their creation and removed from the simulation. The effects of neutron transport and re-interactions are therefore excluded by this approach. These first generation neutrons are produced almost exclusively by photo-nuclear and muon-nuclear inelastic scattering reactions, referred to as photo-nuclear and muon-nuclear reactions, for both simulation programs. Findings for the first generation neutrons from the Geant4 and FLUKA simulations are presented in Section~\ref{sec:Geant4_Fluka_First_Generation_Neutrons}. The recorded first generation neutrons were then used as input to Geant4 and FLUKA simulations of the MINIDEX Run~2 setup, in the last step of the comparison. The contribution of the first generation neutrons to the neutron signal rate R$_{\text{S}}^{\text{Sim}}$ for Geant4 and FLUKA is discussed in Section~\ref{sec:Geant4_Fluka_Neutron_Signal_Rate_Contribution}.

\subsection{Muon Interactions}
\label{sec:Geant4_Fluka_Muon_Interaction}

\begin{figure} 
\centering \begin{subfigure}[htbp]{16.3cm} \includegraphics[scale=0.75]{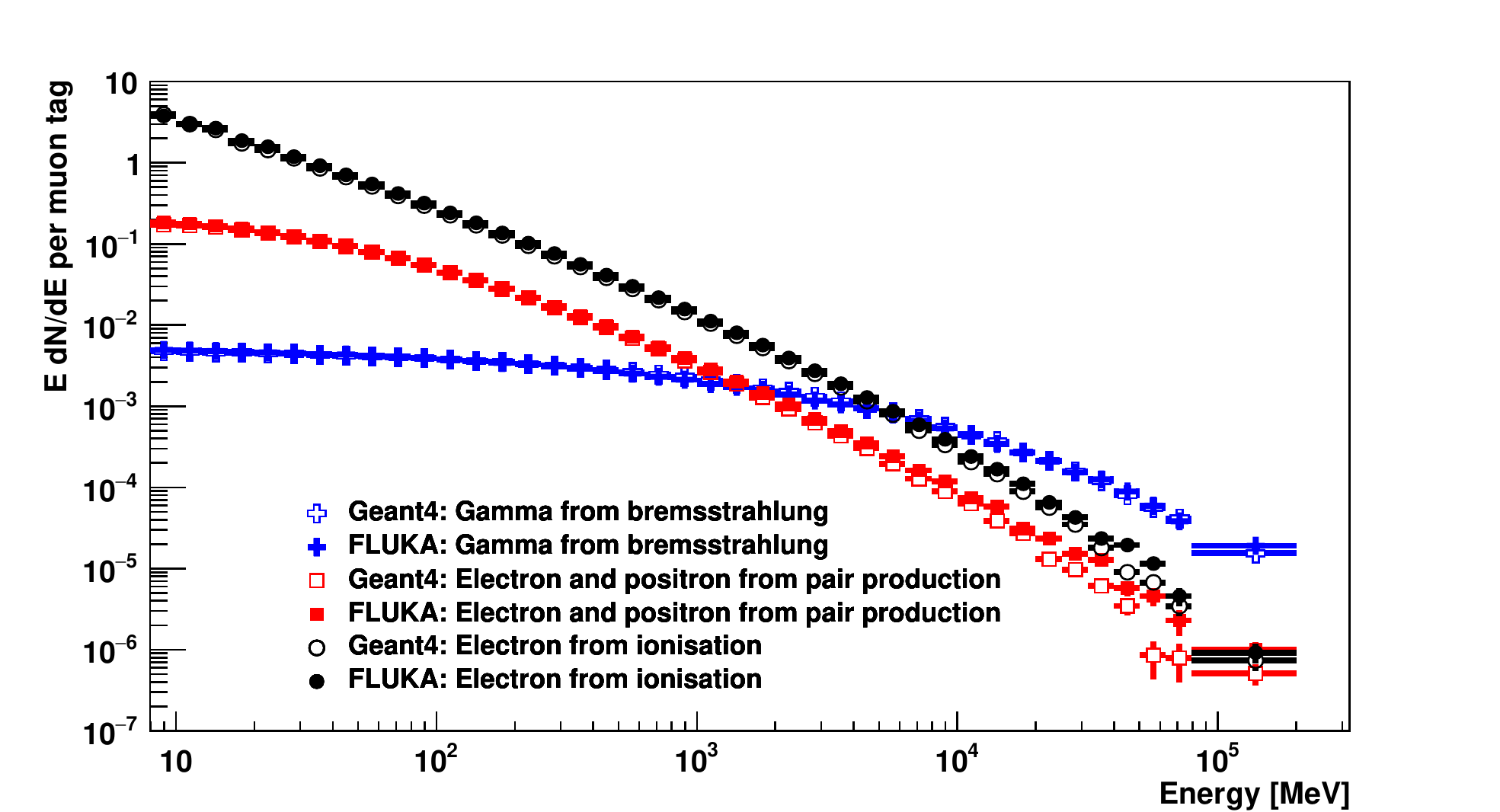}
\centering\caption{}
\end{subfigure}
\centering \begin{subfigure}[htbp]{16.3cm} \includegraphics[scale=0.75]{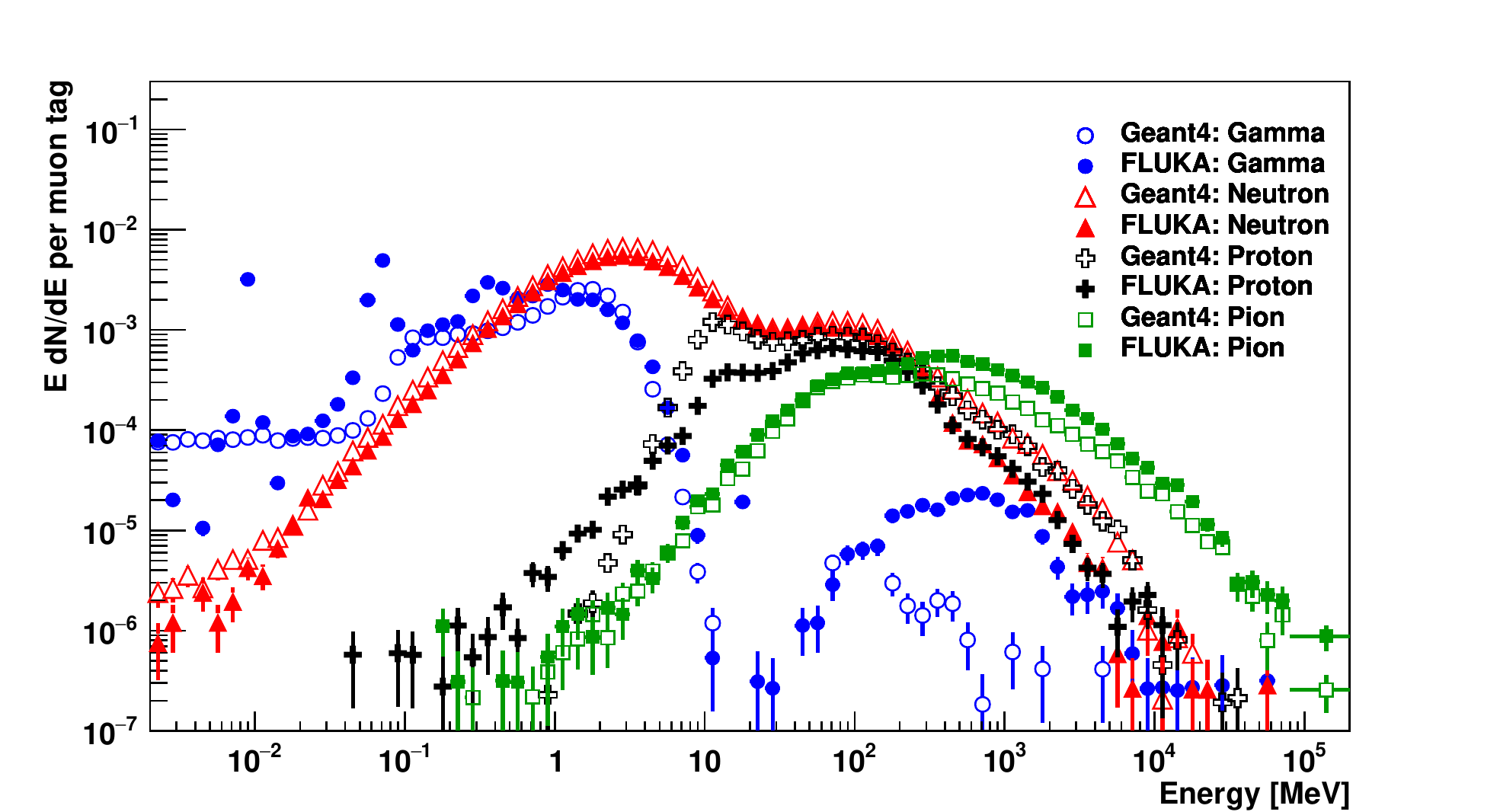}
\centering\caption{}
\end{subfigure}
\caption{Differential energy spectra of secondaries multiplied by energy (E). The secondaries were generated by tagged muons in the lead target walls of the MINIDEX Run~2 setup. The spectra obtained with Geant4 and FLUKA are displayed by open and filled markers, respectively. In (a) the secondaries from bremsstrahlung (blue crosses), electron-positron pair production (electrons and positrons are plotted together, red squares) and ionisation (black circles) are depicted. In (b) gammas (blue circles), neutrons (red triangles), protons (black crosses) and pions (green squares) generated by muon-nuclear reactions are shown. In Table~\ref{tab:muon_secondaries} the production rate of secondaries, generated in muon-nuclear reactions, are given. Assigned uncertainties are statistical only and are partially smaller than the marker size.} 
\label{fig:comparison_MAGE_FLUKA_WithoutRatio}
\end{figure}

At first, only secondaries, defined as particles produced directly by muons inside the lead target walls, were analysed. There are four main processes by which muons lose energy during their passage through matter: bremsstrahlung, electron-positron pair production, ionisation (i.e. muon-electron scattering) and muon-nuclear reactions. The energy spectra of secondaries generated by muons for the first three processes, i.e. electrons, positrons and photons, are depicted in Fig.~\ref{fig:comparison_MAGE_FLUKA_WithoutRatio}(a)\footnote{Only electrons, positrons and gammas with energies $>$\,8\,MeV are shown. The reason is that if the energy of the secondaries falls below $\approx$\,8\,MeV only a negligible number of neutrons are produced within lead. This energy threshold for the neutron production was determined independently for Geant4 and FLUKA.}.
Note, most plots throughout Section~\ref{sec:mage_fluka_comparison} show lethargy~\cite{international1970iaea}. Good agreement for the Geant4 and FLUKA predicted distributions is found with only small deviations for secondaries at energies above a few GeV. In Fig.~\ref{fig:comparison_MAGE_FLUKA_WithoutRatio}(b) the energy spectra of the most frequently produced secondaries for muon-nuclear reactions are compared. The main features of the distributions are similar while for energies above a few hundred MeV there are significant deviations in the predictions for the production rates of the individual secondaries. In Table~\ref{tab:muon_secondaries} the production rates of secondaries, generated in muon-nuclear reactions, for energies above and below 20\,MeV are given for both simulation tools.
\begin{table}[ht]
\centering
\caption{Most frequently produced secondaries predicted by Geant4 and FLUKA in muon-nuclear reactions within the lead target walls of MINIDEX Run~2. The rates for Geant4 and FLUKA are expressed in number of 10$^{-5}$ secondaries per muon tag. In addition the ratios between the FLUKA and Geant4 predicted rate of secondaries are given. The stated uncertainties are statistical only.}
\begin{tabular}{llll}
 \hline
 & \textbf{Geant4} & \textbf{FLUKA}  & \textbf{Ratio}    \\
& \multicolumn{2}{c}{[10$^{-5}$] per muon tag} &  \\
\hline
Gamma $\leq$\,20\,MeV & 577.0\,$\pm$\,1.6 & 964.5\,$\pm$\,2.5 & 1.67\,$\pm$\,0.01 \\
Gamma $>$\,20\,MeV & 0.39\,$\pm$\,0.04 & 5.30\,$\pm$\,0.19 & 13.6\,$\pm$\,1.5 \\
\hline
Neutron $\leq$\,20\,MeV & 1697.9\,$\pm$\,2.8 & 1418.1\,$\pm$\,3.0 & 0.84\,$\pm$\,0.01 \\
Neutron $>$\,20\,MeV & 306.9\,$\pm$\,1.2  & 265.8\,$\pm$\,1.3 & 0.87\,$\pm$\,0.01 \\
\hline
Proton $\leq$\,20\,MeV & 109.6\,$\pm$\,0.7 & 31.1\,$\pm$\,0.5 & 0.28\,$\pm$\,0.01 \\
Proton $>$\,20\,MeV & 227.5\,$\pm$\,1.0 & 155.6\,$\pm$\,1.0 & 0.68\,$\pm$\,0.01 \\
\hline
Pion $\leq$\,20\,MeV & 3.13\,$\pm$\,0.12 & 4.19\,$\pm$\,0.17 & 1.34\,$\pm$\,0.07 \\
Pion $>$\,20\,MeV & 129.3\,$\pm$\,0.8 & 182.8\,$\pm$\,1.1 & 1.41\,$\pm$\,0.01 \\
\hline
\end{tabular}
\label{tab:muon_secondaries}
\end{table}
It can be seen that FLUKA and Geant4 predictions differ significantly, depending on the particle type and energy range. This is especially apparent for the predicted rate of gammas with energies $>$\,20\,MeV.

\begin{figure}[htbp] 
\centering
\includegraphics[scale=0.48]{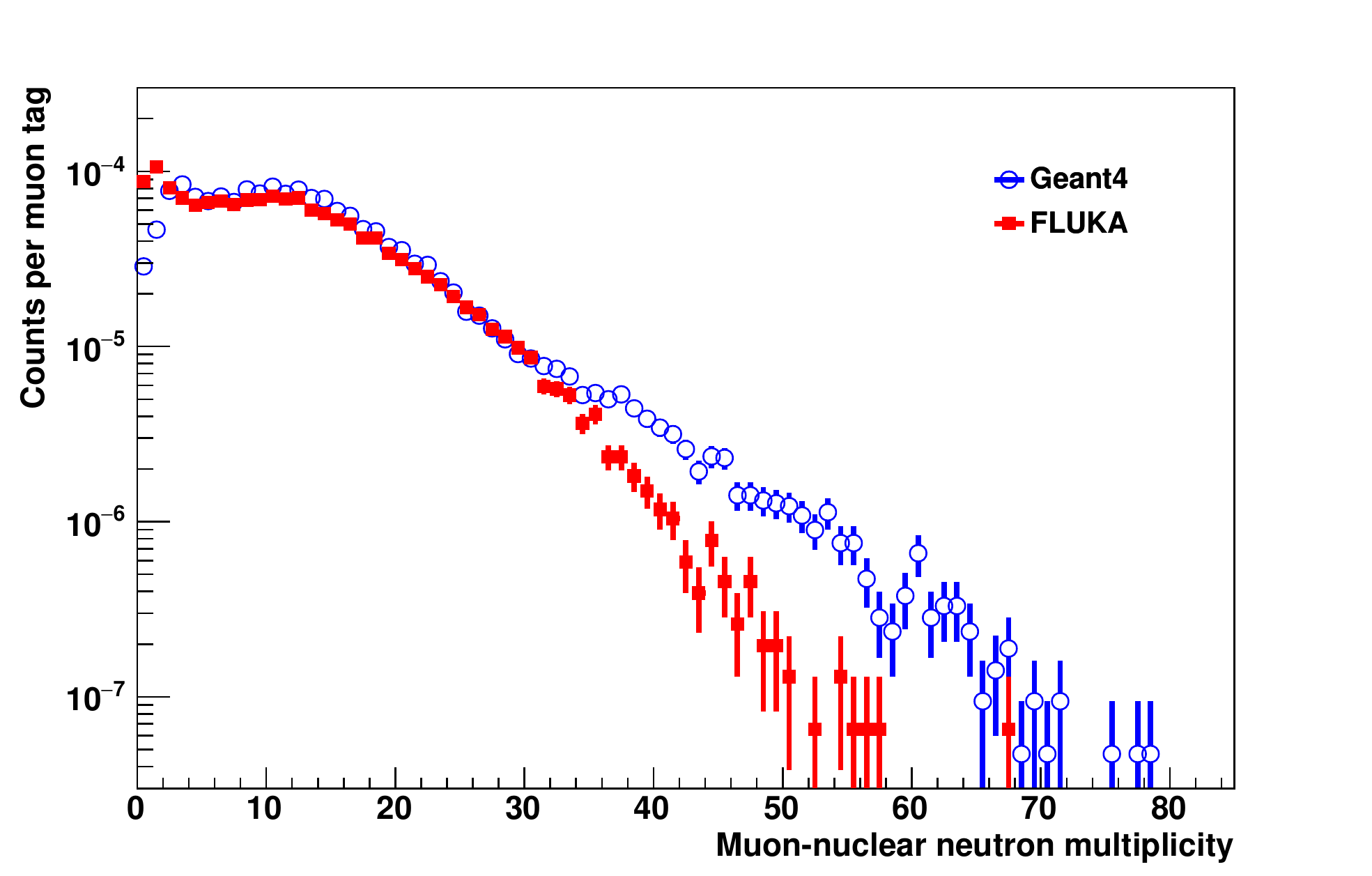}
\caption{Multiplicity of neutrons produced directly by muons in muon-nuclear reactions. Only reactions that occurred in the lead target walls of the MINIDEX Run~2 setup for events with a muon tag are included. The distributions obtained with Geant4 and FLUKA are represented by blue circles and red squares, respectively. The maximum observed number of muon-nuclear reactions within the same muon tag for Geant4 and FLUKA is one. The mean muon-nuclear neutron multiplicity predicted by Geant4 and FLUKA is 13.19\,$\pm$\,0.05 and 10.98\,$\pm$\,0.05, respectively. Assigned uncertainties are statistical only and are partially smaller than the marker size.} 
\label{fig:MuonNuclearNeutronMultiplicity_fig}
\end{figure}

In Fig.~\ref{fig:MuonNuclearNeutronMultiplicity_fig} the multiplicity of neutrons produced in muon-nuclear reactions (muon-nuclear neutron multiplicity) is shown.
A mean muon-nuclear neutron multiplicity of 13.19\,$\pm$\,0.05 and 10.98\,$\pm$\,0.05 was found for Geant4 and FLUKA, respectively. The total rate of muon-nuclear reactions in Geant4 and FLUKA was determined to be (158.0\,$\pm$\,0.9)\,$\cdot$\,10$^{-5}$ and (153.3\,$\pm$\,1.0)\,$\cdot$\,10$^{-5}$, respectively. While the total rate is similar, large discrepancies are found especially for muon-nuclear reactions with high neutron multiplicities. Note that for the used version of FLUKA the rate of muon-nuclear reactions with low neutron multiplicities is likely to be underpredicted~\cite{empl2015fluka}.

\subsection{First Generation Neutron Production}
\label{sec:Geant4_Fluka_First_Generation_Neutrons}

\begin{figure}[htbp] 
\centering
\includegraphics[scale=0.75]{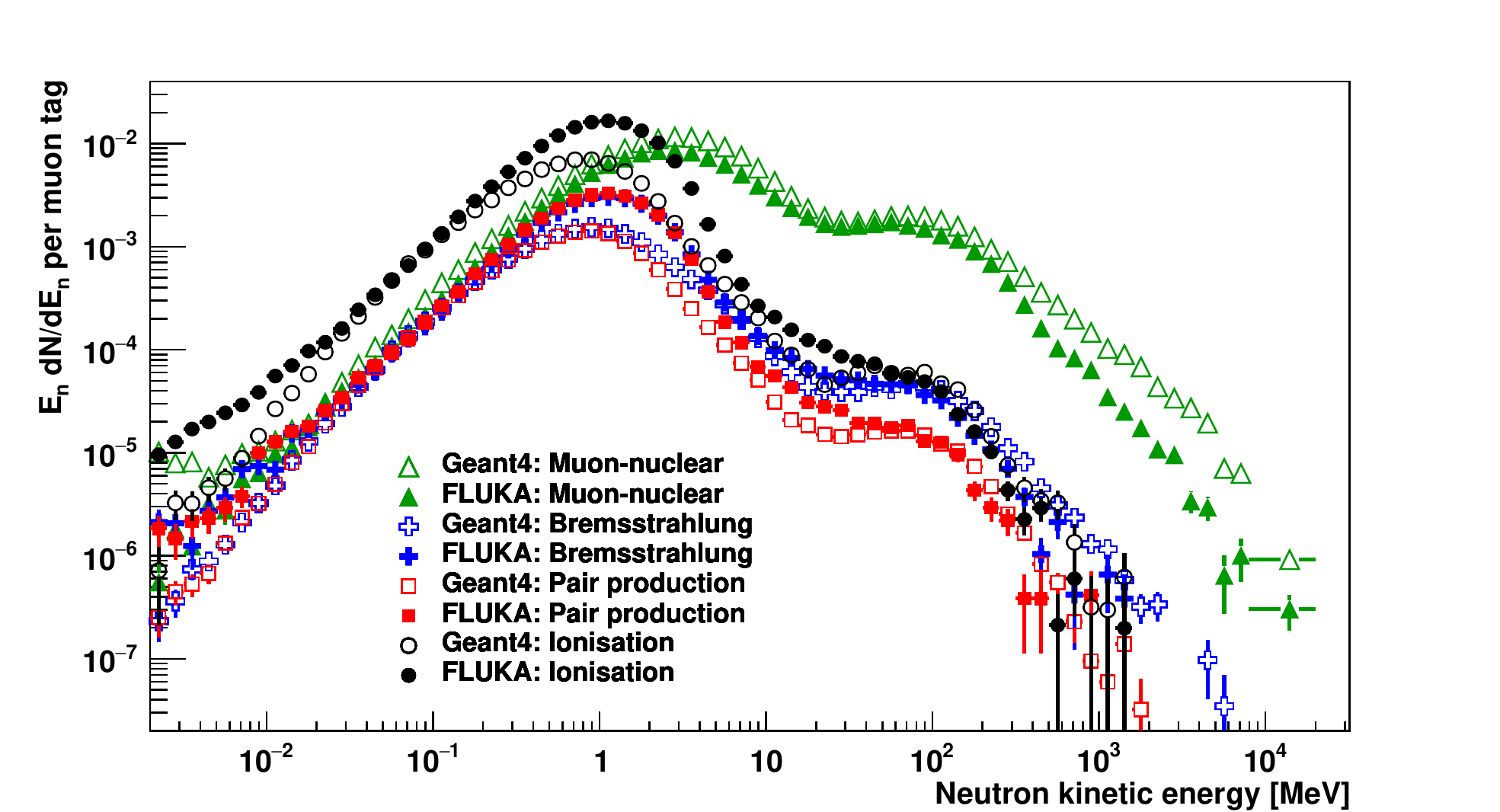}
\caption{Differential kinetic energy spectra of first generation neutrons multiplied by kinetic energy (E$_{\text{n}}$). Only first generation neutrons that were generated in the lead target walls of the MINIDEX Run~2 setup are included. The kinetic energy spectra predicted by Geant4 and FLUKA are displayed by open and filled markers, respectively. The first generation neutrons for the four investigated muon interaction processes are displayed. Assigned uncertainties are statistical only and are partially smaller than the marker size. In Table~\ref{tab:first_generation_neutrons} the rate of first generation neutrons for the four muon interactions processes are presented.} 
\label{fig:comparison_MAGE_FLUKA_WithoutRatio_SecondStep}
\end{figure}

In a second step, all muon-induced neutrons were recorded at the position of their creation and subsequently removed from the simulation. Hence, neutron transport and neutron re-interactions were excluded by this approach. These neutrons, referred to as first generation neutrons, were either produced directly by a muon (i.e. in muon-nuclear reactions) or indirectly by any muon-induced particle (except neutrons). Only neutrons that were generated in the lead target walls were selected and compared between Geant4 and FLUKA. In Fig.~\ref{fig:comparison_MAGE_FLUKA_WithoutRatio_SecondStep} the resulting energy spectra of the first generation neutrons of the four investigated muon interactions are depicted.
For Geant4 the indirectly produced neutrons account for (42.44\,$\pm$\,0.12)\,$\%$ of the total rate of first generation neutrons from muon-nuclear reactions whereas for FLUKA a value of (36.24\,$\pm$\,0.15)\,$\%$ was found. It is evident from Fig.~\ref{fig:comparison_MAGE_FLUKA_WithoutRatio_SecondStep} that the mean energy of first generation neutrons predicted by Geant4 and FLUKA for each of the four investigated muon interactions is inconsistent. For muon-nuclear reactions a value of $\approx$\,25\,MeV and $\approx$\,20\,MeV was obtained for Geant4 and FLUKA, respectively, whereas for the other muon interactions deviating values between 1.5\,MeV and 5\,MeV were determined. In Table~\ref{tab:first_generation_neutrons} the rate of first generation neutrons for energies above and below 20\,MeV are presented for Geant4 and FLUKA for the different muon interactions.
\begin{table}[ht]
\centering
\caption{Rate of first generation neutrons predicted by Geant4 and FLUKA for the four investigated muon interaction processes in the lead target walls of MINIDEX Run~2. The rates for Geant4 and FLUKA are expressed in number of 10$^{-5}$ first generation neutrons per muon tag. In addition, the ratios between the FLUKA and Geant4 predictions are given. The stated uncertainties are statistical only.}
\begin{tabular}{llll}
 \hline
 & \textbf{Geant4}  & \textbf{FLUKA}  & \textbf{Ratio}    \\
& \multicolumn{2}{c}{[10$^{-5}$] per muon tag} &  \\
\hline
Muon-nuclear $\leq$\,20\,MeV & 2991\,$\pm$\,5 & 2256.4\,$\pm$\,3.3 & 0.75\,$\pm$\,0.01 \\
Muon-nuclear $>$\,20\,MeV & 491.8\,$\pm$\,1.9 & 384.4\,$\pm$\,1.4 &0.78\,$\pm$\,0.01 \\
\hline
Bremsstrahlung $\leq$\,20\,MeV & 382.2\,$\pm$\,1.7 & 649.4\,$\pm$\,1.8 & 1.70\,$\pm$\,0.01 \\
Bremsstrahlung $>$\,20\,MeV & 10.13\,$\pm$\,0.28 & 9.86\,$\pm$\,0.22 & 0.97\,$\pm$\,0.03 \\
\hline
Pair production $\leq$\,20\,MeV & 324.7\,$\pm$\,1.6 & 677.5\,$\pm$\,1.8 & 2.09\,$\pm$\,0.01 \\
Pair production $>$\,20\,MeV & 3.43\,$\pm$\,0.16 & 4.03\,$\pm$\,0.14 & 1.17\,$\pm$\,0.07 \\
\hline
Ionisation $\leq$\,20\,MeV & 1578.9\,$\pm$\,3.5 & 3403\,$\pm$\,4 & 2.16\,$\pm$\,0.01 \\
Ionisation $>$\,20\,MeV & 12.81\,$\pm$\,0.31 & 14.00\,$\pm$\,0.26 & 1.09\,$\pm$\,0.03 \\
\hline
\end{tabular}
\label{tab:first_generation_neutrons}
\end{table}
The rate of first generation neutrons predicted by Geant4 for muon-nuclear reactions is (31.89\,$\pm$\,0.26)\,$\%$ higher than the corresponding rate from FLUKA. The Geant4 predicted combined rate of first generation neutrons from the other three muon interaction processes is smaller than the FLUKA predicted rate by (51.41\,$\pm$\,0.10)\,$\%$.

Furthermore, by studying the processes leading to the production of first generation neutrons in the case of ionisation, pair production and bremsstrahlung, it was found that for Geant4 and FLUKA $>$\,94\,$\%$ of these neutrons are produced in photo-nuclear reactions. This means that in the case of ionisation and pair production the electrons and positrons in general do not generate neutrons directly. Typically the electrons and positrons lead to the generation of photons which then interact inelastically with nuclei and produce neutrons.

\subsection{Neutron Signal Rate Contribution}
\label{sec:Geant4_Fluka_Neutron_Signal_Rate_Contribution}

Finally, all first generation neutrons recorded in the previous step were simulated in Geant4 and FLUKA in order to determine the individual contributions of the four investigated muon interactions to the predicted neutron signal rate of the full simulation R$_{\text{S}}^{\text{Sim}}$. In this case neutron transport and neutron re-interactions are considered. In Table~\ref{tab:first_generation_neutrons_contribution} the resulting contributions are given together with R$_{\text{S}}^{\text{Sim}}$.
\begin{table}[ht]
\centering
\caption{Contributions of the first generation neutrons in lead from the different muon interaction processes to the neutron signal rate of the full simulation R$_{\text{S}}^{\text{Sim}}$ in Geant4 and FLUKA. All presented rates are expressed in number of 10$^{-5}$ neutron signals per muon tag. The given uncertainties are statistical only.}
\begin{tabular}{lll}
\hline
   & \textbf{Geant4} & \textbf{FLUKA}      \\
    & \multicolumn{2}{c}{[10$^{-5}$] per muon tag} \\
\hline
  R$_{\text{S}}^{\text{Sim}}$ & 3.37\,$\pm$\,0.09 &  3.60\,$\pm$\,0.13 \\
\hline
First generation neutrons & 2.198\,$\pm$\,0.013 & 2.393\,$\pm$\,0.021  \\
\hline
  Muon-nuclear & 1.588\,$\pm$\,0.011   & 1.093\,$\pm$\,0.014  \\
  Bremsstrahlung & 0.109\,$\pm$\,0.001   & 0.194\,$\pm$\,0.006  \\
  Pair production & 0.086\,$\pm$\,0.001   & 0.187\,$\pm$\,0.006  \\
  Ionisation & 0.415\,$\pm$\,0.006 & 0.919\,$\pm$\,0.013  \\
\hline
\end{tabular}
\label{tab:first_generation_neutrons_contribution}
\end{table}
The sum of detected 2.2\,MeV gammas resulting from all first generation neutrons corresponds to (65.2\,$\pm$\,1.8)\,$\%$ and (66.5\,$\pm$\,2.5)\,$\%$ of R$_{\text{S}}^{\text{Sim}}$ for Geant4 and FLUKA, respectively. The remaining approximately one third of 2.2\,MeV gammas result for example from neutrons produced as a consequence of muon interactions outside the target walls. While the fractions agree well, the contributions of the first generation neutrons from the individual muon interaction processes do not agree. In Geant4 the first generation neutrons from muon-nuclear reactions lead to (47.1\,$\pm$\,1.3)\,$\%$ of R$_{\text{S}}^{\text{Sim}}$ while the first generation neutrons from the other three muon interaction processes contribute only (18.1\,$\pm$\,0.5)\,$\%$. In comparison, the first generation neutrons from muon-nuclear reactions in FLUKA lead to (30.4\,$\pm$\,1.2)\,$\%$ of R$_{\text{S}}^{\text{Sim}}$, while the first generation neutrons from the other three muon interaction processes make up for (36.1\,$\pm$\,1.4)\,$\%$.

\subsection{Conclusion of Geant4 and FLUKA Comparison Study}

In~\cite{Quantification_validity_phonton_nuclear} muon-induced neutrons, predominantly produced in tungsten by photo-nuclear reactions, were measured and the experimental findings were compared to Geant4 and FLUKA predictions. While the flux and the energy spectrum of emitted neutrons from the target was well reproduced by FLUKA, the Geant4 predictions deviated significantly. Geant4 underestimated the flux of emitted neutrons and at the same time did not correctly predict their energies. Furthermore, the Geant4 and FLUKA implemented cross sections for the neutron production in photo-nuclear reactions in natural tungsten, lead and zinc were compared to the cross sections recommended by the International Atomic Energy Agency (IAEA)~\cite{iaea_website}. While the FLUKA implemented cross sections agree well with the IAEA recommended ones, large deviations were found for the Geant4 implemented cross sections. In~\cite{Luo2016} the Geant4 implemented cross sections for the neutron production in photo-nuclear reactions for various materials (e.g. $^{133}$Cs and $^{197}$Au) were compared to the cross sections provided by the Exchange Format experimental nuclear reaction database~\cite{exfor}. Large discrepancies between the Geant4 implemented cross sections and the ones from the experimental data base were reported. This provides further evidence for an inaccurate treatment of the neutron production in photo-nuclear reactions in Geant4.

The presented Geant4 and FLUKA simulation study together with the briefly introduced publications and the agreement of FLUKA with the measurement indicate that Geant4, when used with Geant4 recommended or standard physics lists\footnote{The description of photo-nuclear reactions in Geant4, when used with the recommended or standard physics lists, is always the same~\cite{geant4_photo_nuclear}. This is also valid for muon-nuclear reactions~\cite{geant4_muon_nuclear}. Furthermore, when MaGe simulations are carried out with the Default or the Shielding physics list, the description of photo-nuclear and muon-nuclear reactions is identical to the ones of Geant4~\cite{Boswell:2010mr}.}, significantly underpredicts the neutron production in lead by photo-nuclear reactions for muon energies at shallow depths. Since Geant4 at the same time does reproduce the measured neutron signal rate R$_{\text{S}}^{\text{Exp}}$, it may imply that the neutron production in muon-nuclear reactions is significantly overpredicted by Geant4. However, an inaccurate treatment of the transport and interactions of hadrons by Geant4, which depends on the chosen physics list, could also contribute.

\section{Conclusion and Outlook}

In order to understand the background from muon-induced neutrons in current and future low-background experiments, reliable MC simulations are crucial. To evaluate and tune simulation tools in the context of muon-induced neutrons, experimental data sets are essential. MINIDEX, with its successful measurement of muon-induced neutrons for lead and copper high-Z target materials, provides such data sets for muons with a mean energy of (8.7\,$\pm$\,0.2)\,GeV and (8.5\,$\pm$\,0.2)\,GeV, respectively. The comparison of results from the analysis of experimental data and Geant4 generated MC data for copper revealed a discrepancy in the number of detected 2.2\,MeV gammas (from muon-induced neutrons captured on hydrogen) per muon tag. The measured value is overpredicted by Geant4 by (39\,$\pm$\,28)\,$\%$. On the other hand, the results from the analysis of the experimental and Geant4 generated MC data for lead showed good agreement for the number of detected 2.2\,MeV gammas per muon tag. At the same time, the predicted rate for lead found with a complementary FLUKA simulation is in good agreement with the measured value as well.

A detailed study of muon interactions and neutron production in lead by Geant4 and FLUKA was carried out. Discrepancies in the energy and rate of neutrons, generated in muon-nuclear and photo-nuclear reactions in lead, were found. Together with recent publications~\cite{Quantification_validity_phonton_nuclear,Luo2016} this indicates that Geant4, when used with Geant4 recommended or standard physics lists, does not correctly describe photo-nuclear and muon-nuclear inelastic scattering reactions in lead for muon energies at shallow underground sites. The observed underpredicted neutron production in photo-nuclear reactions in Geant4 for lead seems to be compensated for by an overprediction of the neutron production in muon-nuclear reactions. As a consequence, the predictions of Geant4 in the context of muon-induced neutrons in high-Z materials, especially at shallow underground sites, should be treated with caution. The use of alternative simulation programs, such as FLUKA, is suggested to cross check Geant4 predictions concerning the neutron production in muon-nuclear and photo-nuclear reactions. In general, whenever particles from muon-nuclear or photo-nuclear reactions constitute a non-negligible background, careful evaluation of simulation predictions is recommended.

Currently, muon-induced neutrons produced by muon capture on lead are measured and investigated~\cite{Oliver_Plaul_master_thesis}. The near future plans of the MINIDEX project are to study muon-induced neutrons generated in aluminum and iron from through-going and stopping muons simultaneously at the same underground site. In the long term it would be possible to measure further materials (e.g. marble or concrete) or to move to a different underground site. A further experimental location would provide the possibility to measure muon-induced neutrons at a different mean muon energy.

\section*{Acknowledgements}

We thank the groups of Peter Grabmayr and Josef Jochum from the University of T{\"u}bingen for providing us help and space in their underground laboratory. We especially thank Igor Usherov for his relentless support and hospitality over the last years. We also want to thank Anna Zsigmond for her valuable and helpful comments to this manuscript.

\section*{}

\bibliography{mybibfile}

\begin{thebibliography}{10}
\expandafter\ifx\csname url\endcsname\relax
  \def\url#1{\texttt{#1}}\fi
\expandafter\ifx\csname urlprefix\endcsname\relax\def\urlprefix{URL }\fi
\expandafter\ifx\csname href\endcsname\relax
  \def\href#1#2{#2} \def\path#1{#1}\fi

\bibitem{PhysRevD.86.010001}
J.~Beringer, et~al., Review of particle physics, Phys. Rev. D 86 (2012) 010001.
\newblock \href {http://dx.doi.org/10.1103/PhysRevD.86.010001}
  {\path{doi:10.1103/PhysRevD.86.010001}}.

\bibitem{gerda_active_shielding}
M.~Agostini, et~al., {Searching Neutrinoless Double Beta Decay with Gerda Phase
  II}, Int. J. Mod. Phys 46 (2018) 1860040.
\newblock \href {http://dx.doi.org/10.1142/S2010194518600406}
  {\path{doi:10.1142/S2010194518600406}}.

\bibitem{Heusser}
G.~Heusser, {Low-Radioactivity Background Techniques}, Annu. Rev. Nucl. Part.
  Sci. 45~(1) (1995) 543--590.
\newblock \href {http://dx.doi.org/10.1146/annurev.ns.45.120195.002551}
  {\path{doi:10.1146/annurev.ns.45.120195.002551}}.

\bibitem{Agostinelli2003250}
S.~Agostinelli, et~al., {Geant4 — a simulation toolkit}, Nucl. Inst. Meth. A
  506~(3) (2003) 250 -- 303.
\newblock \href {http://dx.doi.org/10.1016/S0168-9002(03)01368-8}
  {\path{doi:10.1016/S0168-9002(03)01368-8}}.

\bibitem{Reichhart:2013xkd}
L.~Reichhart, et~al., {Measurement and simulation of the muon-induced neutron
  yield in lead}, Astropart. Phys. 47 (2013) 67--76.
\newblock \href {http://dx.doi.org/10.1016/j.astropartphys.2013.06.002}
  {\path{doi:10.1016/j.astropartphys.2013.06.002}}.

\bibitem{Kluck:2013xga}
H.~M. Kluck, {Measurement of the Cosmic-Induced Neutron Yield at the Modane
  Underground Laboratory}, Ph.D. thesis, KIT, Karlsruhe (2013).
\newblock \href {http://dx.doi.org/10.1007/978-3-319-18527-9}
  {\path{doi:10.1007/978-3-319-18527-9}}.

\bibitem{matteos_paper}
I.~Abt, et~al., {The Muon-Induced Neutron Indirect Detection EXperiment,
  MINIDEX}, Astropart. Phys. 90 (2017) 1 -- 13.
\newblock \href {http://dx.doi.org/10.1016/j.astropartphys.2017.01.011}
  {\path{doi:10.1016/j.astropartphys.2017.01.011}}.

\bibitem{Boswell:2010mr}
M.~Boswell, et~al., {MaGe - a Geant4-based Monte Carlo Application Framework
  for Low-background Germanium Experiments}, IEEE Trans. Nucl. Sci. 58 (2011)
  1212--1220.
\newblock \href {http://dx.doi.org/10.1109/TNS.2011.2144619}
  {\path{doi:10.1109/TNS.2011.2144619}}.

\bibitem{Battistoni2015}
G.~Battistoni, et~al., {Overview of the FLUKA code}, Ann. Nucl. Energy 82
  (2015) 10 -- 18.
\newblock \href {http://dx.doi.org/10.1016/j.anucene.2014.11.007}
  {\path{doi:10.1016/j.anucene.2014.11.007}}.

\bibitem{ferrari2005fluka}
A.~Ferrari, et~al., {FLUKA: A multi-particle transport code (Program version
  2005)}.

\bibitem{promt_gamma_database}
{International Atomic Energy Agency},
  \href{http://www-pub.iaea.org/MTCD/publications/PDF/Pub1263_web.pdf}{{Database
  of prompt gamma rays from slow neutron capture for elemental analysis}}
  (2007).
\newline\urlprefix\url{http://www-pub.iaea.org/MTCD/publications/PDF/Pub1263_web.pdf}

\bibitem{Raphael_phd}
R.~Knei{\ss}l, {Ph.D. thesis, in progress} (publishing expected 2019).

\bibitem{Data_sheet_bc408}
I.~Saint-Gobain Ceramics \&~Plastics,
  \href{https://www.crystals.saint-gobain.com/sites/imdf.crystals.com/files/documents/sgc-bc400-404-408-412-416-data-sheet.pdf}{{Premium
  Plastic Scintillators}}, data sheet (Aug 2016).
\newline\urlprefix\url{https://www.crystals.saint-gobain.com/sites/imdf.crystals.com/files/documents/sgc-bc400-404-408-412-416-data-sheet.pdf}

\bibitem{Data_sheet_PMT}
{ET Enterprises}, \href{https://my.et-enterprises.com/pdf/9266B.pdf}{{9266B
  series data sheet}}, data sheet (Aug 2010).
\newline\urlprefix\url{https://my.et-enterprises.com/pdf/9266B.pdf}

\bibitem{Data_sheet_germanium}
{Mirion Technologies},
  \href{http://www.canberra.com/products/detectors/pdf/Germanium-Det-SS-C39606.pdf}{{Germanium
  Detectors}}, data sheet (2016).
\newline\urlprefix\url{http://www.canberra.com/products/detectors/pdf/Germanium-Det-SS-C39606.pdf}

\bibitem{Data_sheet_extended_range_germanium_detector}
{Mirion Technologies},
  \href{http://canberra.com/products/detectors/pdf/XtRa-detectors-C49310.pdf}{{Extended
  Range Coaxial Ge Detectors}}, data sheet (Nov 2016).
\newline\urlprefix\url{http://canberra.com/products/detectors/pdf/XtRa-detectors-C49310.pdf}

\bibitem{Struck_ADC}
{Struck Innovative Systeme}, \href{https://www.struck.de/sis3316.html}{{SIS3316
  16 Channel VME Digitizer Family}}, webpage (2018).
\newline\urlprefix\url{https://www.struck.de/sis3316.html}

\bibitem{battistoni_2011}
G.~Battistoni, et~al., {FLUKA as a new high energy cosmic ray generator}, Nucl.
  Instrum. Methods Phys. Res. A 626 - 627 (2011) S191 -- S192.
\newblock \href {http://dx.doi.org/10.1016/j.nima.2010.05.019}
  {\path{doi:10.1016/j.nima.2010.05.019}}.

\bibitem{fritz}
{Fritz, M.}, {Verm\"ogen und Bau Baden-W\"urttemberg, Amt T\"ubingen}, private
  communication.

\bibitem{goerz}
{G\"orz, E.}, {Verm\"ogen und Bau Baden-W\"urttemberg, Amt T\"ubingen}, private
  communication.

\bibitem{Abt2008}
I.~Abt, et~al., {Neutron interactions as seen by a segmented germanium
  detector}, Eur. Phys. J. A 36~(2) (2008) 139--149.
\newblock \href {http://dx.doi.org/10.1140/epja/i2007-10553-8}
  {\path{doi:10.1140/epja/i2007-10553-8}}.

\bibitem{lindote2009simulation}
A.~Lindote, et~al., {Simulation of neutrons produced by high-energy muons
  underground}, Astropart. Phys. 31~(5) (2009) 366--375.
\newblock \href {http://dx.doi.org/10.1016/j.astropartphys.2009.03.008}
  {\path{doi:10.1016/j.astropartphys.2009.03.008}}.

\bibitem{Mail_mirion_technology_unceratainty_size_germanium}
{Mirion Technologies}, {Email to author by H. Krueger} (Sep 2018).

\bibitem{Shielding_Physics_List_webpage}
D.~H. Wright,
  \href{http://www.slac.stanford.edu/comp/physics/geant4/slac_physics_lists/shielding/physlistdoc.html}{{Shielding
  Physics List Description}}, Website (2017-12-07).
\newline\urlprefix\url{http://www.slac.stanford.edu/comp/physics/geant4/slac_physics_lists/shielding/physlistdoc.html}

\bibitem{Quantification_validity_phonton_nuclear}
L.~Quintieri, et~al., {Quantification of the validity of simulations based on
  Geant4 and FLUKA for photo-nuclear interactions in the high energy range},
  EPJ Web Conf. 153 (2017) 06023.
\newblock \href {http://dx.doi.org/10.1051/epjconf/201715306023}
  {\path{doi:10.1051/epjconf/201715306023}}.

\bibitem{Luo2016}
W.~Luo, et~al., {A data-based photonuclear simulation algorithm for determining
  specific activity of medical radioisotopes}, Nucl. Sci. Tech. 27~(5) (2016)
  113.
\newblock \href {http://dx.doi.org/10.1007/s41365-016-0111-9}
  {\path{doi:10.1007/s41365-016-0111-9}}.

\bibitem{international1970iaea}
{International Atomic Energy Agency},
  \href{http://www-pub.iaea.org/books/IAEABooks/1206/Neutron-Fluence-Measurements}{{Neutron
  Fluence Measurements}}, no. 107 in Technical Reports Series, Vienna, 1970.
\newline\urlprefix\url{http://www-pub.iaea.org/books/IAEABooks/1206/Neutron-Fluence-Measurements}

\bibitem{empl2015fluka}
AIP Publishing, FLUKA: Predictive power for cosmogenic backgrounds, Vol. 1672.
\newblock \href {http://dx.doi.org/10.1063/1.4927999}
  {\path{doi:10.1063/1.4927999}}.

\bibitem{iaea_website}
{International Atomic Energy Agency - Nuclear Data Services},
  \url{https://www-nds.iaea.org}.

\bibitem{exfor}
{Exchange Format Experimental Nuclear Reaction Database},
  \url{https://www-nds.iaea.org/exfor/exfor.htm}.

\bibitem{geant4_photo_nuclear}
{Geant4 Collaboration},
  \href{http://geant4-userdoc.web.cern.ch/geant4-userdoc/UsersGuides/PhysicsReferenceManual/html/photolepton_hadron/gammaNuclear.html}{{Gamma-nuclear
  Interactions}}, {Physics reference manual, Webpage} (2018).
\newline\urlprefix\url{http://geant4-userdoc.web.cern.ch/geant4-userdoc/UsersGuides/PhysicsReferenceManual/html/photolepton_hadron/gammaNuclear.html}

\bibitem{geant4_muon_nuclear}
{Geant4 Collaboration},
  \href{http://geant4-userdoc.web.cern.ch/geant4-userdoc/UsersGuides/PhysicsReferenceManual/html/photolepton_hadron/muonNuclear.html}{{Muon-nuclear
  Interactions}}, {Physics reference manual, Webpage} (2018).
\newline\urlprefix\url{http://geant4-userdoc.web.cern.ch/geant4-userdoc/UsersGuides/PhysicsReferenceManual/html/photolepton_hadron/muonNuclear.html}

\bibitem{Oliver_Plaul_master_thesis}
O.~Plaul, {Measurement of Neutrons from Muon Capture on Lead with MINIDEX},
  Master's thesis, Technical University Munich (2018).

\end{thebibliography}

\end{document}